\documentclass[a4paper,11pt,reqno]{amsart}
\usepackage{amsmath,amsthm,amssymb}
\usepackage{mathrsfs}
\usepackage{float}
\usepackage{graphicx}
\usepackage{braket}
\allowdisplaybreaks
\numberwithin{equation}{section}
\theoremstyle{definition}
\newtheorem{definition}{Definition}[section]

\newtheorem{example}[definition]{Example}
\newtheorem{theorem}[definition]{Theorem}

\newtheorem{corollary}[definition]{Corollary}
\newtheorem{lemma}[definition]{Lemma}

\newtheorem{fact}[definition]{Fact}

\newcommand{\NPb}{{\rlap{\raise.4ex\hbox{$\scriptscriptstyle\bullet$}}
\lower.4ex\hbox{$\scriptscriptstyle\bullet$}}}
\newcommand{\NPc}{{\rlap{\raise.4ex\hbox{$\scriptscriptstyle\circ$}}
\lower.4ex\hbox{$\scriptscriptstyle\circ$}}}
\newcommand{\seteq}{\mathbin{:=}}

\newcommand{\vl}{\vec{\lambda}}
\newcommand{\vm}{\vec{\mu}}

\newcommand{\lo}{\lambda^{(1)}}
\newcommand{\lt}{\lambda^{(2)}}
\newcommand{\lN}{\lambda^{(N)}}
\newcommand{\mo}{\mu^{(1)}}
\newcommand{\mt}{\mu^{(2)}}
\newcommand{\mN}{\mu^{(N)}}
\newcommand{\vu}{\vec{u}}
\newcommand{\vv}{\vec{v}}

\newcommand{\Xo}{X^{(1)}}

\newcommand{\vr}{\vec{r}}
\newcommand{\vs}{\vec{s}}
\newcommand{\lrs}{\lambda_{\vr,\vs}}
\newcommand{\bi}{b^{(i)}}
\newcommand{\bj}{b^{(j)}}

\newcommand{\ao}{a^{(1)}}

\newcommand{\ai}{a^{(i)}}
\newcommand{\aN}{a^{(N)}}

\newcommand{\hi}{h^{(i)}}
\newcommand{\hj}{h^{(j)}}

\newcommand{\sa}{\mathsf{a}}

\newcommand{\sai}{\sa^{(i)}}
\newcommand{\saj}{\sa^{(j)}}

\newcommand{\Qi}{Q^{(i)}}
\newcommand{\Qj}{Q^{(j)}}

\newcommand{\lc}{\mathsf{lt}}
\newcommand{\overstar}[1]{\mathop{\overset{*}{ #1 }}}

\title[Kac Determinant and Singular Vector of DIM Algebra]{Kac determinant and singular vector of the level N representation of Ding-Iohara-Miki algebra}
\author{Yusuke Ohkubo}
\date{}
\keywords{Macdonald symmetric function, Ding-Iohara-Miki algebra, AGT correspondence}
\email{yusuke.ohkubo.math@gmail.com}
\address{Faculty of Mathematics, The National Research University 
Higher School of Economics,  Russian Federation}
\subjclass[2010]{81R10, 33D52, 81R50}
\begin{document}
%%%%%%%%%%%%%%%%%%%%%%%%%%%%%%%%%%%%%%%%%%%

%%%%Title%%%%
\maketitle
%%%%%%%%%%%%%

%%%%%%%%%%%%%%%%%%%%%%%%%%%%%%%%%%%%%%%%%%%%%%%%%%%%%%%%%%%%%%%%%%
%%%%%%%%%%%%%%%%%%%%%%%%%%%%%%abstruct%%%%%%%%%%%%%%%%%%%%%%%%%%%%
%%%%%%%%%%%%%%%%%%%%%%%%%%%%%%%%%%%%%%%%%%%%%%%%%%%%%%%%%%%%%%%%%%
\begin{abstract}
In this paper, 
we obtain the formula for the Kac determinant of the algebra 
arising from the level $N$ representation of the Ding-Iohara-Miki algebra. 
It is also discovered that 
its singular vectors 
correspond to generalized Macdonald functions 
(the q-deformed version of the AFLT basis). 
\end{abstract}
%%%%%%%%%%%%%%%%%%%%%%%%%%%%%%%%%%%%%%%%%%%%%%%%%%%%%%%%%%%%%%%%%%
%%%%%%%%%%%%%%%%%%%%%%%%%%%%%%%%%%%%%%%%%%%%%%%%%%%%%%%%%%%%%%%%%%
%%%%%%%%%%%%%%%%%%%%%%%%%%%%%%%%%%%%%%%%%%%%%%%%%%%%%%%%%%%%%%%%%%

%%%%%%%%%%%%%%%%%%%%%%%%%%%%%%%%%%%%%%%%%%%%%%%%%%%%%%%%%%%%%%%%%%
%%%%%%%%%%%%%%%%%%%%%%%%%%Introduction%%%%%%%%%%%%%%%%%%%%%%%%%%%%
%%%%%%%%%%%%%%%%%%%%%%%%%%%%%%%%%%%%%%%%%%%%%%%%%%%%%%%%%%%%%%%%%%
\section{Introduction}

\begin{figure}[H]
\begin{center}
\includegraphics[width=5cm]{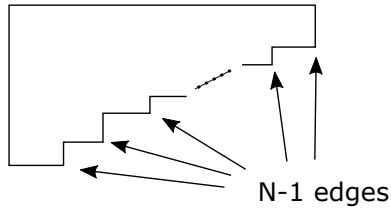}
\caption{Young diagram with $N-1$ edges}
\label{fig:YoungDiag_N-1edges}
\end{center}
\end{figure}

In 1995, Mimachi and Yamada discovered the surprising relation 
that the Jack functions with rectangular Young diagrams 
have a one-to-one correspondence with singular vectors of the Virasoro algebra 
\cite{MimachiYamada:1995}. 
As a $q$-difference deformation of the Jack polynomials, 
there is a system of orthogonal functions 
called the Macdonald functions \cite{Macdonald}. 
Awata, Kubo, Odake and Shiraishi
introduced in \cite{SKAO:1995:quantum}
a $q$-deformation of the Virasoro algebra,
which is named the deformed Virasoro algebra. 
This deformed algebra is designed so that singular vectors 
of Verma modules correspond to Macdonald symmetric functions 
associated with rectangular Young diagrams. 
It is also possible to obtain the Jack and Macodnald functions 
with general partitions  from the singular vectors of the $W_N$-algebra 
and the deformed $W_N$-algebra, 
which is the (deformed) Virasoro algebra when $N = 2$ 
\cite{MimachiYamada:RIMSkokyuroku, AKOS:1995Excited, Awata:1995Quantum}. 
To be exact, 
singular vectors of the (deformed) $W_N$-algebra 
can be realized by $N-1$ families of bosons 
under the free field representation. 
By a certain projection to one of these bosons, 
we can obtain 
the Jack (or Macdonald) functions associated with Young diagrams 
with $N-1$ edges  
(see Figure \ref{fig:YoungDiag_N-1edges}).  
As we will see later in Introduction, 
we discover a sort of generalization of these correspondence 
using generalized Macdonald functions arising from AGT conjecture. 

AGT (Alday-Gaiotto-Tachikawa) conjecture states that 
the Virasoro conformal blocks of two-dimensional conformal field theories
correspond to the instanton partition functions of four-dimensional 
$\mathcal{N}=2$ $SU(2)$ supersymmetric gauge 
theories \cite{alday2010liouville}. 
It is also expected that the four-dimensional gauge theories 
with the higher gauge group $SU(N)$ correspond with the $W_N$-algebra 
\cite{Wyllard:2009}. 
$q$-deformed version of the AGT conjecture is also provided. 
That is, the deformed Virasoro/$W$-algebra is related 
to five-dimensional gauge theories (5D AGT conjecture) 
\cite{AwataYamada1, AwataYamada2}. 
It is known that a good basis called AFLT basis 
(Alba-Fateev-Litvinov-Tarnopolski basis) 
\cite{alba2011combinatorial,FL:2011Integrable} %belavin2011agt} 
exists in the representation space of the tensor product of 
Virasoro/$W_N$ algebra and $U(1)$ Heisenberg algebra, 
and by this basis 
the conformal block can be combinatorially expanded 
like the Nekrasov partition functions. 
Since the AFLT basis corresponds to the torus fixed points 
in the instanton moduli space, 
it is also called the fixed point basis. 
The AFLT basis can be regarded as a sort of generalization of 
Jack functions under the free field representation. 
Also Dotsenko-Fateev integral are expanded 
in the form of the Nekrasov funciton 
with the help of that generalized Jack functions 
\cite{morozov2013finalizing}. 
$q$-deformed version of the AFLT basis \cite{awata2011notes} 
can be constructed in the representation space 
of the Ding-Iohara-Miki algebra $\mathcal{U}$ (DIM algebra), 
and regarded as generalized Macdonald functions. 
By using, the generalized Macdonald fucntions, 
$q$-Dotsenko-Fateev integral are also combinatorially 
expanded \cite{Zenkevich:2014}.

The DIM algebra is the associative algebra generated by four currents 
$x^\pm(z)=\sum_{n\in \mathbb{Z}}x^\pm_n z^{-n}$, 
$\psi^\pm(z)=\sum_{\pm n\in \mathbb{Z}_{\ge0}}\psi^\pm_n z^{-n}$ 
and the central element $\gamma^{\pm 1/2}$ satisfying the relations 
(\ref{eq:def rel of DIM1})-(\ref{eq:def rel of DIM2}) 
as explained in Appendix \ref{sec:Def of DIM}.  
The DIM algebra $\mathcal{U}$  
has the face of a $q$-deformation of the $W_{1+\infty}$ algebra 
as introduced by Miki in \cite{Miki:2007}, 
and the deformed Virasoro/$W$-algebra appear  
in its representation \cite{FHSSY}. 
Since the DIM algebra has a lot of background, 
there are a lot of other names 
such as quantum toroidal $\mathfrak{gl}_1$ algebra 
\cite{FJMM:2015:Quantum, FJMM:2016:Finite}, 
elliptic Hall algebra \cite{BS:2012:I} and so on. 
Furthermore, 
the DIM algebra $\mathcal{U}$ is associated with the Macdonald functions. 
The generator $x^+_0$ can be essentially identified 
with Macdonald's difference operator 
under the free field representation 
$\rho_u : \mathcal{U} \rightarrow \mathrm{End}(\mathcal{F}_{u})$, 
where $\mathcal{F}_u$ is the Fock module and 
$u$ expresses its highest wight \cite{FHHSY}. 
(See also Fact \ref{fact:lv. 1 rep of DIM}.) 
The DIM algebra $\mathcal{U}$ has a Hopf algebra structure 
which does not exist in the deformed Virasoro/$W$-algebra. 
Using the coproduct $\Delta$ of the DIM algebra, 
we can consider the tensor representation of $\rho_u$: 
\begin{align}
\rho^{(N)}_{\vu} &:= \rho_{u_1} \otimes \cdots \otimes \rho_{u_N} \circ 
\Delta^{(N)}, \\%: \mathcal{U} \rightarrow \mathrm{End}(\mathcal{F}_{\vu}), \\
\Delta^{(N)}& := (\Delta \otimes id \otimes \cdots \otimes id ) 
\circ \cdots \circ (\Delta \otimes id) \circ \Delta 
: \mathcal{U} \rightarrow \mathcal{U}^{\otimes N},  
\end{align}
where $\vu=(u_1,\ldots,u_N)$. 
This representation $\rho^{(N)}_{\vu}$ 
is called the level $N$ representation.  
%or the level $(N,0)$ representation. 
The generalized Macdonald functions are defined to be 
eigenfunctions of the generator $X^{(1)}_0$ given by 
\begin{eqnarray}
X^{(1)}(z):=\sum_{n\in \mathbb{Z}} X^{(1)}_n z^{-n} :=\rho^{(N)}(x^+(z)). 
\end{eqnarray}
Whereas the ordinary Macdonald functions are indexed by partitions, 
the generalized Macdonald ones 
are indexed by $N$-tuples of partitions $\vl=(\lo, \ldots , \lN)$ 
and denoted by $\Ket{P_{\vl}}$. 
They correspond to the ordinary Macdonald function in the case of $N=1$. 
Introducing the generator $X^{(i)}_n$ 
which appearing in the commutation relations of $X^{(j)}_n$ ($j=1,\ldots ,i-1 $), 
we can define an algebra $\mathcal{A}(N):=\langle X^{(i)}_n | i=1,\ldots, N ,n\in \mathbb{Z}\rangle$. 
(See Definition \ref{df:x^i_n}.) 
With the help of the integral form of generalized Macdonald functions, 
which are defined by using PBW type vectors 
\begin{align} 
 &\Ket{X_{\vec{\lambda}}} \seteq 
  X^{(1)}_{-\lambda^{(1)}_1} X^{(1)}_{-\lambda^{(1)}_2} \cdots  X^{(2)}_{-\lambda^{(2)}_1} X^{(2)}_{-\lambda^{(2)}_2}\cdots X^{(N)}_{-\lambda^{(N)}_1} X^{(N)}_{-\lambda^{(N)}_2} \cdots \Ket{\vec{u}}, 
\end{align} 
it was conjectured 
that the conformal block of a certain vertex operator of DIM algebra 
corresponds to 5D (K-theoretic) partition functions \cite{awata2011notes}.
The geometric interpretation of this kind of correspondence 
is given in  \cite{Negut:2016:qAGTW}, 
and the algebra $\mathcal{A}(N)$ coincides with  
the $qW$ algebra of type $\mathfrak{gl}_N$ (not $\mathfrak{sl}_N$) 
which is important in that interpretation. 
Let us also mention that by using another representation of DIM algebra, 
it is proved that the correlation function 
of an intertwining operator of DIM algebra 
coincides with the 5D Nekrasov formula \cite{awata2012quantum}. 
Furthermore, 
it is conjectured that R-matrix of DIM algebra can be written in terms of 
the integral form of generalized Macdonald functions 
\cite{AKMMMOZ:2016:Toric,AKMMMOZ:2016:Anomaly,FHMZ:2017:Maulik-Okounkov}

In this paper, 
we focus on the representation of the algebra $\mathcal{A}(N)$. 
One of the results in this paper (Theorem \ref{thm:KacDet}) 
is the formula for 
the Kac determinant with respect to the PBW type vector 
$\Ket{X_{\vec{\lambda}}}$.

%\begin{thm*}

\noindent 
\textbf{Theorem.}
\begin{align}\label{eq:Kac det in Intro}
&\det\left( \Braket{X_{\vl}|X_{\vm}} \right)_{|\vl|=|\vm|=n}=
\prod_{\vl \vdash n} \prod_{k=1}^N b_{\lambda^{(k)}}(q) b'_{\lambda^{(k)}}(t^{-1})
%\times (u_1 u_2 \cdots u_N)^{2\sum_{\vl \vdash n} \ell(\lambda^{(N)})}
\\
& \quad \times \prod_{\substack{1\leq r,s\\ rs\leq n}}
\left( 
(u_1 u_2 \cdots u_N)^{2}
 \prod_{1\leq i < j \leq N} (u_i-q^st^{-r}u_j)(u_i-q^{-r}t^s u_j) \right)^{P^{(N)}(n-rs)}, \nonumber
\end{align}
where 
$b_{\lambda}(q) \seteq \prod_{i\geq 1} \prod_{k=1}^{m_i} (1-q^k)$, 
$b'_{\lambda}(q) \seteq \prod_{i\geq 1} \prod_{k=1}^{m_i} (-1+q^k)$, and   
$P^{(N)}(n)$ 
denotes the number of the $N$-tuples of Young diagrams of size $n$. 
Further, 
$m_i=m_i(\lambda)$ is the number of entries in $\lambda$ equal to $i$. 
%\end{thm*}

This determinant can be proved 
by using the fact that the generators $X^{(i)}_n$ 
can be decomposed into the deformed $W$-algebra part and the $U(1)$ part 
by a linear transformation of the bosons, 
and using the screening currents of the deformed $W$-algebra. 
By this formula, 
we can solve the conjecture \cite[Conjecture 3.4]{awata2011notes} that 
the PBW type vectors $\Ket{X_{\vec{\lambda}}}$ 
of the algebra $\mathcal{A}(N)$  
are a basis.

We also discover that singular vectors of the algebra 
$\mathcal{A}(N)$ 
correspond to the generalized Macdonald functions 
(Theorem \ref{thm:Sing vct and Gn Mac 2}). 
By this result, 
we can get singular vectors from generalized Macdonald functions. 
The singular vectors are intrinsically 
the same as those of the deformed $W$-algebra. 
However, 
as the projection of bosons is required 
for coincidence with the ordinary Macdonald functions 
\cite{Awata:1995Quantum}, 
the result of this paper that does not need projections 
can be regarded as an extension of 
correspondence with ordinary Macdonald functions.  
As a corollary of this fact, 
we can find a new relation of the ordinary Macdonald functions 
and the generalized Macdonald functions 
by the projection of bosons. 
%Furthermore, since screening operators are written by integrals, 
%we can also get an integral representation of generalized Macdonald  %functions. 

\begin{figure}[H]
\begin{center}
\includegraphics[width=12cm]{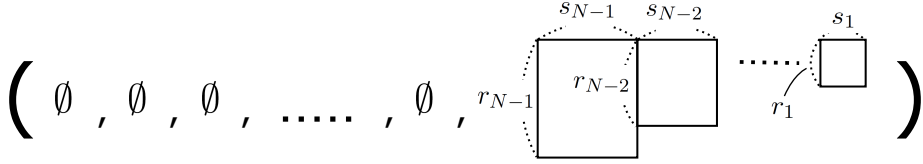}
\caption{Young diagram corresponding to singular vector 
in the case that $r_{k+1} \geq r_k$ for all $k$. }
\label{fig:YoungDiag_OnlyRightSide}
\end{center}
\end{figure}

Concretely, 
the vector $\Ket{\chi_{\vec{r}, \vec{s}}}$ defined to be 
\begin{align}
\Ket{\chi_{\vec{r}, \vec{s}}} \seteq  
\oint \prod_{k=1}^{N-1} \prod_{i=1}^{r_k} dz^{(k)}_i 
&S^{(N-1)}(z^{(N-1)}_1) \cdots S^{(N-1)}(z^{(N-1)}_{r_{N-1}}) \cdots \nonumber \\
&\cdots S^{(1)}(z^{(1)}_1) \cdots S^{(1)}(z^{(1)}_{r_{1}}) 
\ket{\vv}
\end{align}
is a singular vector. 
Here $S^{(i)}(z)$ denotes the screening operator defined 
in (\ref{eq:def of screening}), 
and $\vv=(v_1, \ldots , v_N)$ is an $N$-tuple of 
parameters depending on the number of screening currents $r_k$ and 
positive integers $s_k$ ($k=1,\ldots, N-1$). 
For more details, see Section \ref{sec:sing vct and Gn Mac}. 
In the case that $r_k$ satisfy $0 \leq r_k \leq r_{k+1}$ for all $k$, 
the singular vector $\Ket{\chi_{\vec{r}, \vec{s}}}$ 
coincides with the generalized Macdonald function 
$\Ket{P_{\vl}}$ with the $N$-tuple of 
Young diagrams in Figure \ref{fig:YoungDiag_OnlyRightSide}. 
In fact, Figure \ref{fig:YoungDiag_N-1edges} means 
the same Young diagram being on the rightmost side in Figure 
\ref{fig:YoungDiag_OnlyRightSide}. 
Hence the projection of this generalized Macdonald function 
corresponds to the ordinary Macdonald functions 
associated with the rightmost Young diagram with $N-1$ edges in Figure 
\ref{fig:YoungDiag_OnlyRightSide} 
(Corollary \ref{cor:projection of Gn Mac}). 
When the condition $r_{k+1} \geq r_{k}$ is removed, 
the above figure is not a Young diagram. 
However it turns out that 
the vector $\Ket{\chi_{\vec{r}, \vec{s}}}$ coincides with 
the generalized Macdonald function 
obtained by cutting off the protruding part 
and moving boxes to the Young diagrams on the left side.
For details, 
see Theorem \ref{thm:Sing vct and Gn Mac 2} and 
Example \ref{ex:Gn Mac and sing. vct 1}.

This paper is organized as follows. 
In Section \ref{sec:lv N rep alg}, 
we define the algebra $\mathcal{A}(N)= \langle X^{(i)}_n \rangle$ 
and give a factorized formula for the Kac determinant. 
Section \ref{sec:Proof of Kacdet} is devoted to 
the proof of that formula. 
Along \cite{FHSSY}, 
we provide the linear transformations of bosons 
to decompose the algebra $\mathcal{A}(N)$ 
into the deformed $W$-algebra and $U(1)$ Heisenberg algebra. 
By using some results of the deformed $W$-algebra, 
we prove the formula for the Kac determinant. 
In Section \ref{sec:sing vct and Gn Mac}, 
we give the definition of the generalized Macdonald functions 
and show that the singular vectors $\ket{\chi_{\vr,\vs}}$ 
coincide with generalized Macdonald functions. 
In Appendix \ref{sec: Macdonald and HL}, 
the definition of ordinary Macdonald functions are explained.  
At last, 
in Appendix \ref{sec:Def of DIM}, 
we describe the definition of the DIM algebra and the level $N$ representation.

\subsection*{Notation of partitions}

Let us explain the notation of partitions and Young diagrams. 
A partition $\lambda=(\lambda_1,\ldots,\lambda_n)$ 
is a non-increasing sequence of integers 
$\lambda_1 \geq \ldots \geq \lambda_n \geq 0$. We write 
$\left| \lambda \right| \seteq \sum_{i} \lambda_i$. 
The length of $\lambda$, denoted by $\ell(\lambda)$, 
is the number of elements $\lambda_i$ with $\lambda_i \neq 0$. 
%Partitions are identified if all elements except $0$ are the same.   
%For example, $(3,2)=(3,2,0)$. 
$m_i=m_i(\lambda)$ denotes the number of elements that are equal to $i$ in $\lambda$, 
and we occasionally write partitions as 
$\lambda=(1^{m_1},2^{m_2},3^{m_3}, \ldots)$.
For example, $\lambda=(6,6,6,2,2,1)=(6^3,2^2,1)$. 
The partitions are identified with the Young diagrams, 
which are the figures written by 
putting  $\lambda_i$ boxes on the $i$-th row and aligning the left side. 
We use $(i,j) \in \mathbb{N}^2$ for a coordinate 
of a box in the $i$-th row and the $j$-th column in a Young diagram. 
For example, 
if $\lambda=(6,4,3,3,1)$, its Young diagram is 
\setlength{\unitlength}{1mm}
\newsavebox{\yang}
\savebox{\yang}{
\begin{picture}(30,25)(0,0)
\put(0,25){\line(1,0){30}}
\put(0,20){\line(1,0){30}}
\put(0,15){\line(1,0){20}}
\put(0,10){\line(1,0){15}}
\put(0,5){\line(1,0){15}}
\put(0,0){\line(1,0){5}}
\put(30,25){\line(0,-1){5}}
\put(25,25){\line(0,-1){5}}
\put(20,25){\line(0,-1){10}}
\put(15,25){\line(0,-1){20}}
\put(10,25){\line(0,-1){20}}
\put(5,25){\line(0,-1){25}}
\put(0,25){\line(0,-1){25}}
\put(11.5,16.5){$s$}
\end{picture}}
\begin{center}
\usebox{\yang}.
\end{center}
The coordinate of the box $s$ is $(2,3)$. 
For an $N$-tuple of partition $\vl=(\lo, \ldots, \lN)$, 
we write $|\vl|:=|\lo|+\cdots+|\lN|$. 
If $|\vl|=m$, 
we occasionally use the symbol $\vdash$ as $\vl \vdash m$.

%%%%%%%%%%%%%%%%%%%%%%%%%%%%%%%%%%%%%%%%%%%%%%%%%%%%%%%%%%%%%%%%%%
%%%%%%%%%%%%%%%%%%%%%%%%%%%%%%%%%%%%%%%%%%%%%%%%%%%%%%%%%%%%%%%%%%
%%%%%%%%%%%%%%%%%%%%%%%%%%%%%%%%%%%%%%%%%%%%%%%%%%%%%%%%%%%%%%%%%%

%%%%%%%%%%%%%%%%%%%%%%%%%%%%%%%%%%%%%%%%%%%%%%%%%%%%%%%%%%%%%%%%%%
%%%%%%%%%%%%%%%%%%%%level N rep of DIM alg%%%%%%%%%%%%%%%%%%%%%%%%
%%%%%%%%%%%%%%%%%%%%%%%%%%%%%%%%%%%%%%%%%%%%%%%%%%%%%%%%%%%%%%%%%%
\section{Kac determinant of level $N$ representation of DIM algebra}
\label{sec:lv N rep alg}

Firstly, we define the algebra $\mathcal{A}(N)=\langle X^{(i)}_n \rangle$ 
obtained by the level $N$ representation of the DIM algebra. 
This algebra $\mathcal{A}(N)$ and the level $N$ representation 
play an important role 
in 5D AGT conjecture corresponding to $SU(N)$ gauge theory. 
Let $q$ and $t=q^{\beta}$ be generic complex parameters 
and $p\seteq q/t$. 
$a_n^{(i)}$ and $Q^{(i)}$ ($n \in \mathbb{Z}$, $i=1,\ldots N$) are 
the generators of the Heisenberg algebra such that 
\begin{equation}
[a^{(i)}_n, a^{(j)}_m]
=n\frac{1-q^{|n|}}{1-t^{|n|}} \delta_{n+m,0} \delta_{i,j}, 
\qquad [a^{(i)}_n,Q^{(j)}]=\delta_{n,0}\delta_{i,j}. 
\end{equation}
Let us define the vertex operators $\eta^{(i)}(z)$ and $\varphi^{(i)}(z)$. 
\begin{definition}Set
\begin{align}
\eta^{(i)} (z)
&:= \exp \left( \sum_{n=1}^{\infty} \frac{1-t^{-n}}{n}\, z^{n} a^{(i)}_{-n} \right) 
   \exp \left( -\sum_{n=1}^{\infty}\frac{(1-t^n)}{n} \, z^{-n} a^{(i)}_n \right), \\
\varphi^{(i)} (z)
&:= \exp \left(  \sum_{n=1}^{\infty} \frac{1-t^{-n}}{n} (1-p^{-n})  z^{n} a^{(i)}_{-n} \right). 
\end{align}
\end{definition}

\begin{definition} \label{df:x^i_n}
Define generators $X^{(i)}(z) = \sum_{n} X^{(i)}_n z^{-n}$ by 
\begin{align}
& X^{(i)}(z) 
 \seteq \sum_{1\leq j_1 <\cdots <j_i \leq N} \NPb \Lambda_{j_1}(z) \cdots \Lambda_{j_i}(p^{i-1}z) \NPb , \\
& \Lambda^{i}(z) 
:= \varphi^{(1)}(z) \varphi^{(2)}(z p^{-\frac{1}{2}}) \cdots \varphi^{(i-1)}(z p^{-\frac{i-2}{2}}) \eta^{(i)}(z p^{-\frac{i-1}{2}}) U_i,
\end{align}
where $\NPb \quad \NPb$ denotes the usual normal ordering product and 
\begin{equation}
U_i:= q^{\sqrt{\beta} a^{(i)}_0} p^{-\frac{N+1}{2}+i}. 
\end{equation} 
\end{definition}

The generator $X^{(1)}(z)$ arises 
from the level $N$ representation of Ding-Iohara-Miki algebra \cite{FHHSY, FHSSY} 
and is obtained by acting the coproduct $N$ times 
to a generator of the DIM algebra corresponding to $\eta(z)$ 
(see Appendix \ref{sec:Def of DIM}). 
The other generators $X^{(i)}_n$ appear in the commutation relations 
of generators $X^{(i-k)}_n$ ($k=1, \ldots i-1$). 
Let $\Ket{0}$ be the highest weight vector in the Fock module 
of the Heisenberg algebra such that $a^{(i)}_n \Ket{0}=0$ for $n \geq0$. 
For an $N$-tuple of complex parameters $\vu=(u_1,\ldots ,u_N)$ 
with $u_i = q^{\sqrt{\beta}w_i}p^{-\frac{N+1}{2}+i}$,  
set $\Ket{\vu}:= e^{\sum_{i=1}^N w_i Q^{(i)}} \Ket{0}$. 
Then they satisfy the relation $U_i \Ket{\vu}= u_i \Ket{\vu}$. 
Similarly, 
let $\Bra{0}$ be the dual highest weight vector, and 
$\Bra{\vu}:=\Bra{0}e^{-\sum_{i=1}^N w_i Q^{(i)}}$. 
$\mathcal{F}_{\vec{u}}$ is the highest weight module generated by 
$\Ket{\vec{u}}$, 
and $\mathcal{F}_{\vec{u}}^*$ is the dual module generated by $\Bra{\vec{u}}$. 
The bilinear form (Shapovalov form) 
$\mathcal{F}_{\vec{u}}^* \otimes \mathcal{F}_{\vec{u}} \rightarrow \mathbb{C}$
is uniquely determined by the condition $\Braket{\vu|\vu}=1$.

\begin{definition}
Define the algebra $\mathcal{A}(N)$ to be the subalgebra 
$\langle X^{(i)}_n | i=1,\ldots, N ,n\in \mathbb{Z}\rangle$ 
in some completion of the endomorphism algebra  
of the Fock module $\mathcal{F}_{\vu}$ for our Heisenberg algebra. 
\end{definition}

\begin{example}\label{ex:comm rel of algebra A}
If $N=2$, 
the commutation relations of the generators are 
\begin{align}
&f^{(1)} \left( \frac{w}{z} \right) X^{(1)}(z) X^{(1)}(w) - X^{(1)}(w) X^{(1)}(z) f^{(1)} \left( \frac{z}{w} \right) \label{eq:rel. of generator X^1} \\ 
& \qquad \qquad \qquad = \frac{(1-q)(1-t^{-1})}{1-p} \left\{ \delta \left( \frac{w}{pz} \right) X^{(2)}(z ) -\delta \left( \frac{pw}{z} \right) X^{(2)}(w ) \right\} , \nonumber \\
&f^{(2)} \left( \frac{w}{z} \right) X^{(2)}(z) X^{(2)}(w) - X^{(2)}(w) X^{(2)}(z) f^{(2)} \left( \frac{z}{w} \right) =0,  \\
&f^{(1)} \left( \frac{pw}{z} \right) X^{(1)}(z) X^{(2)}(w) - X^{(2)}(w) X^{(1)}(z) f^{(1)} \left( \frac{z}{w} \right) =0,
\end{align}
where $\delta(x)=\sum_{n \in \mathbb{Z}} x^n$ is 
the multiplicative delta function 
and  the structure constant $ f^{(i)}(z) =\sum_{l =0}^{\infty} f^{(i)}_l z^{l}$ is defined by 
\begin{equation}
 f^{(1)}(z) \seteq 
 \exp \left\{ \sum_{n >0} \frac{(1-q^n)(1-t^{-n})}{n}z^n \right\}, 
\end{equation}
\begin{equation}
 f^{(2)}(z) \seteq 
 \exp \left\{ \sum_{n >0} \frac{(1-q^n)(1-t^{-n})(1+p^{n})}{n}z^n \right\}. 
\end{equation}
These relations are equivalent to 
\begin{align}
& [X^{(1)}_n, X^{(1)}_m]= -\sum_{l =1}^{\infty} f^{(1)}_l (X^{(1)}_{n-l} X^{(1)}_{m+l} - X^{(1)}_{m-l} X^{(1)}_{n+l}) \nonumber \\
&\qquad \qquad \qquad +\frac{(1-q)(1-t^{-1})}{1-p}(p^m-p^n) X^{(2)}_{n+m},  \\
& [X^{(2)}_n, X^{(2)}_m]= -\sum_{l =1}^{\infty} f^{(2)}_l (X^{(2)}_{n-l} X^{(2)}_{m+l} - X^{(2)}_{m-l} X^{(2)}_{n+l}), \\
& [X^{(1)}_n, X^{(2)}_m]= -\sum_{l =1}^{\infty} f^{(1)}_l (p^{l} X^{(1)}_{n-l} X^{(2)}_{m+l} -  X^{(2)}_{m-l} X^{(1)}_{n+l}) .
\end{align}
\end{example}

The calculation of Example \ref{ex:comm rel of algebra A}
is similar to the ordinary deformed Virasoro and W algebra \cite{Awata:1995Quantum,SKAO:1995:quantum}. 
The algebra $\mathcal{A}(N)$ corresponds to the $qW$ algebra of type 
$\mathfrak{gl}_N$ (not $\mathfrak{sl}_N$) 
defined in \cite{Negut:2016:qAGTW}. 
The relations of generators $X^{(i)}(z)$ 
for the general $N$ case are also shown in \cite{Negut:2016:qAGTW}. 
By the following PBW type vectors $\Ket{X_{\vl}}$, 
we can define the integral form of the generalized Macdonald functions, 
which have significant properties in the AGT correspondence \cite{awata2011notes}.

\begin{definition}\label{df:ordinary PBW vct}
For an $N$-tuple of partitions 
$\vl=(\lo, \lt, \ldots, \lambda^{(N)})$, 
set
\begin{align}
 &\Ket{X_{\vec{\lambda}}} \seteq 
  X^{(1)}_{-\lambda^{(1)}_1} X^{(1)}_{-\lambda^{(1)}_2} \cdots  X^{(2)}_{-\lambda^{(2)}_1} X^{(2)}_{-\lambda^{(2)}_2}\cdots X^{(N)}_{-\lambda^{(N)}_1} X^{(N)}_{-\lambda^{(N)}_2} \cdots \Ket{\vec{u}}, \\
 & \Bra{X_{\vl}} \seteq 
\Bra{\vu} \cdots X^{(N)}_{\lambda^{(N)}_2} X^{(N)}_{\lambda^{(N)}_1} \cdots X^{(2)}_{\lambda^{(2)}_2} X^{(2)}_{\lambda^{(2)}_1} \cdots 
X^{(1)}_{\lambda^{(1)}_2} X^{(1)}_{\lambda^{(1)}_1} . 
\end{align}
\end{definition}

The PBW theorem cannot be used 
because the algebra $\mathcal{A}(N)$ is not a Lie algebra. 
In \cite{awata2011notes} 
it was conjectured that the PBW type vectors $\Ket{X_{\vec{\lambda}}}$ and 
$\Bra{X_{\vec{\lambda}}}$ are a basis over $\mathcal{F}_{\vu}$ 
and $\mathcal{F}_{\vec{u}}^*$, respectively. 
This conjecture can be solved 
by the following Kac determinant of the algebra $\mathcal{A}(N)$.

\begin{theorem}\label{thm:KacDet}
Let $\mathrm{det}_n \seteq 
\det \left( \Braket{X_{\vl}|X_{\vm}} \right)_{\vl, \vm \vdash n}$. 
%Here ordering of vectors $\Ket{X_{\vl}}$ is usual way...
Then  
\begin{align}\label{eq:KacDet for DIM}
\mathrm{det}_n&=
\prod_{\vl \vdash n} \prod_{k=1}^N b_{\lambda^{(k)}}(q) b'_{\lambda^{(k)}}(t^{-1})
%\times (u_1 u_2 \cdots u_N)^{2\sum_{\vl \vdash n} \ell(\lambda^{(N)})}
\\
& \times \prod_{\substack{1\leq r,s\\ rs\leq n}}
\left( 
(u_1 u_2 \cdots u_N)^{2}
 \prod_{1\leq i < j \leq N} (u_i-q^st^{-r}u_j)(u_i-q^{-r}t^s u_j) \right)^{P^{(N)}(n-rs)}, 
\end{align}
where  
$b_{\lambda}(q) \seteq \prod_{i\geq 1} \prod_{k=1}^{m_i} (1-q^k)$, 
$b'_{\lambda}(q) \seteq \prod_{i\geq 1} \prod_{k=1}^{m_i} (-1+q^k)$. 
$P^{(N)}(n)$
denotes the number of $N$-tuples of Young diagrams of size $n$, i.e.,
$\# \big\{ \vl=(\lo, \ldots , \lN) \big| \vl \vdash n\big\}$. 
In particular, if $N=1$, 
\begin{equation}
\mathrm{det}_n=\prod_{\lambda \vdash n} 
b_{\lambda}(q) b'_{\lambda}(t^{-1})
\times u_1^{2\sum_{\lambda \vdash n}\ell(\lambda)}. 
\end{equation}
\end{theorem}

\begin{corollary}
If $u_i \neq 0$ and $u_i \neq q^rt^{-s} u_j$ for any numbers $i$, $j$ and 
integers $r$, $s$, 
then the PBW type vectors $\Ket{X_{\vl}}$ (resp. $\Bra{X_{\vl}}$) are a basis 
over $\mathcal{F}_{\vu}$ (resp. $\mathcal{F}^*_{\vu}$). 
\end{corollary}

Also it can be seen that 
the representation of the algebra $\mathcal{A}(N)$ 
on the Fock Module $\mathcal{F}_{\vu}$ is irreducible 
if and only if the parameters $\vu$ satisfy the condition 
that $u_i \neq 0$ and $u_i \neq q^rt^{-s} u_j$. 
The proof of Theorem \ref{thm:KacDet} is given in the next section.

%%%%%%%%%%%%%%%%%%%%%%%%%%%%%%%%%%%%%%%%%%%%%%%%%%%%%%%%%%%%%%%%%%
%%%%%%%%%%%%%%%%%%%%%%%%%%%%%%%%%%%%%%%%%%%%%%%%%%%%%%%%%%%%%%%%%%
%%%%%%%%%%%%%%%%%%%%%%%%%%%%%%%%%%%%%%%%%%%%%%%%%%%%%%%%%%%%%%%%%%

%%%%%%%%%%%%%%%%%%%%%%%%%%%%%%%%%%%%%%%%%%%%%%%%%%%%%%%%%%%%%%%%%%
%%%%%%%%%%%%%%%%%%Proof of Theorem　thm:KacDet%%%%%%%%%%%%%%%%%%%%
%%%%%%%%%%%%%%%%%%%%%%%%%%%%%%%%%%%%%%%%%%%%%%%%%%%%%%%%%%%%%%%%%%
\section{Proof of Theorem \ref{thm:KacDet}}
\label{sec:Proof of Kacdet}

It is known that 
the algebra $\mathcal{A}(N)$ obtained by 
the level $N$ representation of the DIM algebra 
can be regarded as the tensor product of the deformed $W_N$-algebra and 
the Heisenberg algebra associated with the $U(1)$ factor \cite{FHSSY}. 
This fact is obtained by a linear transformation of bosons. 
The point of proof of Theorem \ref{thm:KacDet} is 
to construct singular vectors by using screening currents 
of the deformed $W_N$-algebra 
under the decomposition of the generators $X^{(i)}(z)$ into 
the deformed $W_N$-algebra part and the $U(1)$ part. 
In general, a vector $\Ket{\chi}$ in the Fock module $\mathcal{F}_{\vu}$ 
is called the singular vector of the algebra $\mathcal{A}(N)$
if it satisfies 
\begin{equation}
X^{(i)}_n \Ket{\chi}=0
\end{equation}
for all $i$ and $n>0$. 
The singular vectors obtained by the screening currents are 
intrinsically the same one of the deformed $W$-algebra. 
From this singular vector, 
we can get the vanishing line of the Kac determinant 
in the similar way of the deformed $W_N$-algebra. 
The formulas for the Kac determinant of 
the deformed Virasoro algebra and the deformed $W_N$-algebra are proved in 
\cite{BouwknegtPilch:1998:Virasoro, BouwknegtPilch:1998:W}.

First, in the $N\geq 2$ case, 
we introduce the following bosons. \\
{\bf U(1) part boson}
\begin{align}
&b'_{-n} \seteq \frac{(1-t^{-n})(1-p^n)}{n(1-p^{Nn})}p^{(N-1)n}\sum_{k=1}^N p^{(\frac{-k+1}{2})n} a^{(k)}_{-n}, \\
&b'_{n} \seteq -\frac{(1-t^{n})(1-p^n)}{n(1-p^{Nn})}p^{(N-1)n}\sum_{k=1}^N p^{\left( \frac{-k+1}{2} \right) n} a^{(k)}_{n}  
\quad (n>0), 
\end{align}
\begin{equation}
b'_{0}\seteq \ao_0 +\cdots +\aN_0, \qquad Q' \seteq \frac{Q^{(1)}+\cdots + Q^{(N)} }{N}.
\end{equation}
\\
{\bf Orthogonal component of $a^{(i)}_n$ for $b'$}
\begin{equation}
b^{(i)}_{-n} \seteq \frac{1-t^{-n}}{n} a^{(i)}_{-n}-p^{(\frac{-i+1}{2})n} b'_{-n}, \quad
b^{(i)}_{n} \seteq \frac{1-t^n}{n}a^{(i)}_{n}+p^{(\frac{-i+1}{2})n} b'_{n} 
\quad (n>0).
\end{equation}
\\
{\bf Fundamental boson of the deformed $W_N$-algebra part}
\begin{equation}
h^{(i)}_{-n}\seteq (1-p^{-n}) \left( \sum_{k=1}^{i-1} p^{\frac{-k+1}{2}n} b^{(k)}_{-n}\right)+  p^{\frac{-i+1}{2}n} b^{(i)}_{-n}, \qquad
h^{(i)}_{n}\seteq -p^{\frac{i-1}{2}n} b^{(i)}_n \quad (n>0),
\end{equation}
\begin{equation}
\hi_0\seteq \ai_0 -\frac{b'_0}{N}, \qquad Q^{(i)}_h\seteq Q^{(i)} - Q'.
\end{equation}
Then they satisfy the following relations
\begin{align}
&[b'_n,b'_{m}]=-\frac{(1-q^{|n|})(1-t^{-|n|})(1-p^{|n|})}{n(1-p^{N|n|})}
\delta_{n+m,0}, \\
&[\bi_n , b'_m]=[\hi_n,b'_m]=[\Qi_h,b'_m]=0, \\
&[\bi_n,\bj_{m}]=\frac{(1-q^{|n|})(1-t^{-|n|})}{n}\delta_{i,j}\delta_{n+m,0} \nonumber \\
&\qquad \qquad \qquad -p^{(N-\frac{i+j}{2})|n|}\frac{(1-q^{|n|})(1-t^{-|n|})(1-p^{|n|})}{n(1-p^{N|n|})}\delta_{n+m,0}, \\
&[\hi_n,\hj_{m}]= -\frac{(1-q^n)(1-t^{-n})(1-p^{(\delta_{i,j}N-1)n})}{n(1-p^{Nn})}p^{Nn \,\theta(i>j)} \delta_{n+m,0}, \\
&[\hi_0,Q^{(j)}_h]=\delta_{i,j}-\frac{1}{N}, \qquad 
\sum_{i=1}^N p^{-in} \hi_n =0, \qquad 
\sum_{i=1}^N  Q^{(i)}_h=0.
\end{align}
where $\theta(P)$ is $1$ or $0$ if the proposition $P$ is true or false,  respectively.%
\footnote{Note that $\hi_n$ and $Q_{h}^{(i)}$ correspond to the fundamental bosons $h^{N-i+1}_n$ and $Q_{h}^{N-i+1}$ in \cite{Awata:1995Quantum}, respectively.} 
Using these bosons, 
we can decompose the generator $X^{(i)}(z)$ into 
the U(1) part and the deformed $W$-algebra part. 
That is to say, 
\begin{equation}
\Lambda_i(z)=\Lambda'_i(z) \Lambda''(z)
\end{equation}
\begin{align}
&\Lambda'_i(z) \seteq \NPb \exp\left( \sum_{n \in \mathbb{Z}_{\neq 0}} \hi_n z^{-n} \right) \NPb q^{\sqrt{\beta} \hi_0} p^{-\frac{N+1}{2}+i}, \\  
&\Lambda''(z) \seteq \NPb  \exp\left( \sum_{n \in \mathbb{Z}_{\neq 0}} b'_{-n}z^n \right) \NPb
q^{\sqrt{\beta} \frac{b'_0}{N}}, 
\end{align}
and
\begin{equation}
X^{(i)}(z) 
= W^{(i)}(z) Y^{(i)}(z)
\end{equation}
\begin{align}
&W^{(i)}(z) 
\seteq \sum_{1\leq j_1 <\cdots <j_i \leq N} 
\NPb \Lambda'_{j_1}(z) \cdots \Lambda'_{j_i}(p^{i-1}z) \NPb, \\
&Y^{(i)}(z) \seteq \NPb \exp\left( \sum_{n\in \mathbb{Z}_{\neq 0}} 
\frac{1-p^{i n}}{1-p^n} b'_{-n}z^n \right) \NPb
q^{\sqrt{\beta}\frac{i b'_0}{N}}. 
\end{align}
$W^{(i)}(z)$ is the generator of the deformed $W_N$-algebra. 
Let us introduce the new parameters $u'_i$ and $u''$ defined by 
\begin{equation}
\prod_{i=1}^N u'_i=1, \quad 
u'' u'_i = u_i \quad (\forall i).
\end{equation}
Then the inner product of PBW type vectors can be written as 
\begin{equation}
\Braket{X_{\vl}|X_{\vm}}=(u'')^{\sum_{k=1}^N k (\ell(\lambda^{(k)})+\ell(\mu^{(k)}))}
\times \left( \mbox{polynomial in $u_1', \ldots , u'_N$  } \right). 
\end{equation}
Hence, its determinant is also in the form 
\begin{equation}
\mathrm{det}_n=(u'')^{2\sum_{\vl \vdash n} \sum_{i=1}^N i \ell(\lambda^{(i)})}
\times F(u'_1,\ldots, u'_N), 
\end{equation}
where 
$F(u'_1,\ldots, u'_N)$ is 
a polynomial in $u'_i$ ($i=1,\ldots,N$) which is independent of $u''$. 
Now in \cite{Awata:1995Quantum}, 
the screening currents of the deformed $W_N$-algebra are introduced: 
\begin{equation}\label{eq:def of screening}
S^{(i)}(z) \seteq 
\NPb \exp \left( \sum_{n\neq 0} \frac{\alpha^{(i)}_n }{1-q^n} \right) \NPb 
e^{\sqrt{\beta} Q^{(i)}_{\alpha}} z^{\sqrt{\beta} \alpha^{(i)}_0}, 
\end{equation}
where $\alpha^{(i)}_n$ is the root boson 
defined by $\alpha^{(i)}_n \seteq h^{(i+1)}_n-h^{(i)}_n$ and 
$\Qi_{\alpha} \seteq Q_h^{(i+1)}-Q_h^{(i)}$. 
The bosons $\alpha^{(i)}_n$ and $\Qi_{\alpha}$ commute with $b'_n$, 
and it is known that 
the screening charge $\oint dz S^{(i)}(z)$ commutes with 
the generators $W^{(j)}(z)$. 
Therefore, 
$\oint dz S^{(i)}(z)$ commutes with any generator $X^{(j)}_n$, 
and it can be regarded as the screening charges of the algebra 
$\mathcal{A}(N)$. 
Define parameters $\hi$ and $\alpha'$ 
by $h^{(i)}+ \frac{\alpha'}{N}=w_i$ and $\sum_{i=1}^N\hi=0$. 
We set $\alpha^{(i)}\seteq h^{(i+1)} -h^{(i)}$. 
For $i=1,\ldots ,N-1$, 
we define $Q^{(i)}_{\Lambda}\seteq \sum_{k=i+1}^N Q^{(k)}_h$. 
Then 
$\Ket{\vu}= e^{\sum_{i=1}^{N-1} \alpha^{(i)} Q^{(i)}_{\Lambda} 
+ \alpha' Q'} \Ket{0}$ 
and we can see the relations 
$[\alpha^{(i)}_n, \Qj_{\Lambda}]=\delta_{i,j}$ and 
$\alpha^{(i)}_0\ket{\vu}=\alpha^{(i)}\ket{\vu}$. 
For any number $i=1,\dots , N$, 
the vector arising from the screening current $S^{(i)}(z)$, 
\begin{equation}\label{eq:sing vct chi(i)rs}
\ket{\chi^{(i)}_{r,s}} = 
\oint dz \prod_{k=1}^r S^{(i)}(z_k) \Ket{\vv}, \quad 
v_{i}=q^{s} t^{r} v_{i+1} \quad  
(r, s \in \mathbb{Z}_{>0})
\end{equation}
is a singular vector. 
$\ket{\chi^{(i)}_{r,s}}$ is in the Fock module $\mathcal{F}_{\vu}$ 
with the parameter $\vu$ satisfying $u_k=v_k$ for $k \neq i, i+1$ and 
$u_{i+1}=t^r v_{i+1}$, $u_{i}=t^{-r} v_{i}$. 
The obtained relation $u_i=q^st^{-r} u_{i+1}$ is equivalent to 
the condition that the parameter $\alpha^{(i)}$ defined above satisfies 
$\alpha^{(i)}=\sqrt{\beta}(1+r)-\frac{1}{\sqrt{\beta}}(1+s)$. 
The $P^{(N)}(n-rs)$ vectors obtained by this singular vector 
\begin{eqnarray}
X_{-\vl} \ket{\chi^{(i)}_{r,s}}, \qquad |\vl|=n-rs
\end{eqnarray}
contribute the vanishing point 
$(u'_i - q^st^{-r} u'_{i+1})^{P^{(N)}(n-rs)}$ in the polynomial $F$. 
Similarly to the case of the deformed $W_N$-algebra 
(see \cite{BouwknegtPilch:1998:W}), 
by the $\mathfrak{sl}_N$ Weyl group invariance of 
the eigenvalues of $W^{(i)}_0$,  
the polynomial $F$ has the factor 
$(u'_i - q^{s}t^{-r} u'_{j})^{P^{(N)}(n-rs)}$ 
($\forall i,j$). 
Considering the degree of polynomials $F(u'_1,\ldots , u'_N)$, 
we can see that when $N \geq 2$, 
the Kac determinant is 
\begin{align}
&\mathrm{det}_n
 = g_{N,n}(q,t) 
    \times (u'')^{2\sum_{\vl \vdash n} \sum_{i=1}^N i\,  \ell(\lambda^{(i)})} \nonumber \\
&\qquad \quad \times \prod_{1 \leq i<j\leq N}
    \prod_{\substack{1\leq r,s \\ rs \leq n}}
     \left( (u'_{i}-q^{s} t^{-r}u'_{j})(u'_{i}-q^{-r} t^{s}u'_{j}) \right)^{P^{(N)}(n-rs)} \nonumber \\ 
&= g_{N,n}(q,t) \nonumber \\
& \quad  \times\prod_{\substack{1\leq r,s \\ rs \leq n}} 
    \left( (u_1u_2\cdots u_N)^2 
      \prod_{1 \leq i<j\leq N}
      (u_{i}-q^{s} t^{-r}u_{j})(u_{i}-q^{-r} t^{s}u_{j}) \right)^{P^{(N)}(n-rs)}, \label{eq:Kacdet1}
\end{align}
where $g_{N,n}(q,t)$ is a rational function in parameters $q$ and $t$
and independent of the parameters $u_i$. 
In (\ref{eq:Kacdet1}), 
we used the following Lemma.

\begin{lemma}
Let $f$ be a (complex valued) function on $\mathbb{N}$. 
Then for any $n \in \mathbb{N}$ and $k=1,\ldots, N$, 
\begin{equation}
\prod_{\vl \vdash n} \prod_{i=1}^{\ell(\lambda^{(k)})} f(\lambda_i)
=\prod_{\substack{r,s \geq 1\\ rs \leq n}} f(r)^{P^{(N)}(n-rs)}. 
\end{equation}
\end{lemma}

This Lemma is a fairly straightforward generalization of 
\cite[Lemma 3.2]{BouwknegtPilch:1998:Virasoro}. 
The proof is similar. 
If $N=1$, the Kac determinant $\mathrm{det}_n$ is clearly in the form 
\begin{equation}
\mathrm{det}_n= g_{1,n}(q,t) \times 
(u_1 \cdots u_N)^{2\sum_{\lambda \vdash n} \ell(\lambda)}. 
\end{equation}

Next, 
the prefactor $g_{N,n}(q,t)$ can be computed in general $N$ case 
by introducing another boson 
\begin{equation}
%\sa^{(i)}_{-n} \seteq \frac{1-t^{-n}}{n}p^{\frac{i-1}{2}n} a^{(i)}_{-n}, \quad
\sa^{(i)}_{n} \seteq 
-\frac{1-t^{n}}{n}p^{\left( \frac{-i+1}{2}\right) n} a^{(i)}_{n}, 
\quad n \in \mathbb{Z}. 
\end{equation}
The commutation relation of the boson $\sai_n$ is
\begin{equation}
[\sai_n, \saj_{-n}] = -\frac{(1-t^{-n})(1-q^n)}{n}\delta_{i,j}, \quad n>0.
\end{equation}
Define the matrix $H^{(n,\pm)}_{\vl,\vm}$ by the expansions 
\begin{equation}
\Ket{X_{\vl}}=\sum_{\vm \vdash n} H^{(n,-)}_{\vl,\vm} \sa_{-\vm}\Ket{\vu}, \quad 
\Bra{X_{\vl}}=\sum_{\vm \vdash n} H^{(n,+)}_{\vl,\vm} \Bra{\vu} \sa_{\vm} \quad (\vl \vdash n), 
\end{equation}
where $\sa_{-\vm}$ and $\sa_{\vm}$ are 
\begin{align}
&\sa_{-\vl}
:=\sa^{(1)}_{-\lo_1}\sa^{(1)}_{-\lo_2} \cdots 
\sa^{(2)}_{-\lt_1}\sa^{(2)}_{-\lt_2} \cdots 
\sa^{(N)}_{-\lN_1}\sa^{(N)}_{-\lN_2} \cdots,\\
&\sa_{\vl}
:=\cdots \sa^{(N)}_{\lN_2} \sa^{(N)}_{\lN_1}
\cdots \sa^{(2)}_{\lt_2} \sa^{(2)}_{\lt_1}
\cdots \sa^{(1)}_{\lo_2} \sa^{(1)}_{\lo_1}. 
\end{align}
We write their determinants as 
$H^{(n,\pm)}\seteq \det (H^{(n,\pm)}_{\vl,\vm})$. 
By using these determinants, 
the Kac determinant can be written as
\begin{equation}
\mathrm{det}_n= H^{(n,+)}\, G_n(q,t)\, H^{(n,-)}. 
\end{equation}
Here $G_n(q,t)$ is the determinant of the diagonal matrix 
$(\Bra{\vu}\sa_{\vl}\sa_{-\vm}\Ket{\vu})_{\vl,\vm \vdash n}$. 
This factor is independent of the parameters $u_i$, 
and we have $G_n(q,t)=\prod_{\vl \vdash n} \prod_{k=1}^N b_{\lambda^{(k)}}(q) b'_{\lambda^{(k)}}(t^{-1})$. 
In (\ref{eq:Kacdet1}), 
the factor depending on $u_i$ in $\mathrm{det}_n$ 
was already clarified. 
Hence, we can determine the prefactor $g_{N,n}(q,t)$ 
by computing the leading term  in $H^{(n,+)}\times H^{(n,-)}$. 
That is, 
the prefactor $g_{N,n}(q,t)$ can be written as 
\begin{equation}
g_{N,n}(q,t) = G_n(q,t) \times 
\left( \mbox{coefficient of } \lc(H^{(n,+)}\times H^{(n,-)};u_1,\ldots,u_N) \right),  
\end{equation}
where we introduce the function $\lc(f;u)$ 
which gives the leading term of $f$ as the polynomial in $u$, 
and 
$\lc(f;u_1,\ldots,u_N)\seteq \lc( \cdots \lc(\lc(f;u_1);u_2)\cdots ;u_N)$. 
To calculate this leading term, 
define the operators $A^{(k)}(z)=\sum_{n\in \mathbb{Z}} A^{(k)}_n z^{-n}$,  
$B^{(k)}(z)=\sum_{n\in \mathbb{Z}}B^{(k)}_n z^{-n}$ by 
\begin{align}
&A^{(k)}(z)=\exp \left\{ \sum_{n>0} \left(\sa^{(k)}_{-n}+\sum_{i=1}^{k-1}p^{(k-i)n} \sai_{-n}\right)z^{n} \right\}, \\ 
&B^{(k)}(z)=\exp \left( \sum_{n>0} \sum_{i=1}^k \sai_n z^{-n} \right). 
\end{align}
$\lc(H^{(n,-)};u_1,\ldots,u_N)$ arises from only the operator 
\begin{equation}
\mathcal{L}^{(k)}(z) \seteq 
\NPb \Lambda_1(z)\Lambda_2(pz) \cdots \Lambda_k(p^{k-1z}) \NPb
\end{equation}
in $X^{(k)}(z)$. 
Then 
\begin{equation}
\mathcal{L}^{(k)}(z)
=U_1\cdots U_k  A^{(k)}(z)B^{(k)}(z). 
\end{equation}
Let $L^{(n,-)}_{\vl,\vm}$ 
and $C^{(n,-)}_{\vl,\vm}$ be the matrices given by 
\begin{equation}
\mathcal{L}_{-\vl}\Ket{\vu}=\sum_{\vm \vdash n} L^{(n,-)}_{\vl,\vm}\, A_{-\vm}\Ket{\vu}, \quad 
A_{-\vl}\Ket{\vu}=\sum_{\vm \vdash n} C^{(n,-)}_{\vl,\vm}\, \sa_{-\vm}\Ket{\vu} 
\quad (\vl \vdash n), 
\end{equation}
where $\mathcal{L}_{-\vl}$ and $A_{-\vl}$ are defined in the usual way: 
\begin{align}
& \mathcal{L}_{-\vl}
=\mathcal{L}^{(1)}_{-\lo_1}\mathcal{L}^{(1)}_{-\lo_2} \cdots 
\mathcal{L}^{(2)}_{-\lt_1}\mathcal{L}^{(2)}_{-\lt_2} \cdots 
\mathcal{L}^{(N)}_{-\lN_1}\mathcal{L}^{(N)}_{-\lN_2} \cdots,  \\
& A_{-\vl}
=A^{(1)}_{-\lo_1}A^{(1)}_{-\lo_2} \cdots 
A^{(2)}_{-\lt_1}A^{(2)}_{-\lt_2} \cdots 
A^{(N)}_{-\lN_1}A^{(N)}_{-\lN_2} \cdots.
\end{align}
Then $\lc(H^{(n,-)};u_1,\ldots,u_N)$ is expressed as  
\begin{equation}
\lc(H^{(n,-)};u_1,\ldots,u_N)
=\det(L^{(n,-)}_{\vl,\vm})
 \det (C^{(n,-)}_{\vl,\vm}). 
\end{equation}
Since the matrix $L^{(n,-)}_{\vl,\vm}$ is lower triangular 
with respect to the partial ordering $\overset{**}{>}^{\mathrm{R}}$ 
\footnote{ 
Here the partial orderings $\overset{**}{>}^{\mathrm{R}}$ 
and $\overset{**}{>}^{\mathrm{L}}$ 
are defined as follows: 
\begin{align}
\vl \overset{**}{\geq}^{\mathrm{R}} \vm \quad \overset{\mathrm{def}}{\Leftrightarrow} \quad
& |\vl|=|\vm|, \quad 
\sum_{i=1}^k |\lambda^{(i)}|  \geq  \sum_{i=1}^k |\mu^{(i)}| \quad (\forall k) \\
& \mathrm{or} \quad "(|\lo|,\ldots, |\lN|)=(|\mo|,\ldots, |\mN|)  
\quad \mathrm{and } \quad 
\lambda^{(i)} \geq \mu^{(i)} \quad (\forall i)" ,
\end{align}
\begin{align}
\vl \overset{**}{\geq}^{\mathrm{L}} \vm \quad \overset{\mathrm{def}}{\Leftrightarrow} \quad
& |\vl|=|\vm|, \quad 
\sum_{i=K}^N |\lambda^{(i)}|  \geq  \sum_{i=k}^N |\mu^{(i)}| \quad (\forall k) \\
& \mathrm{or} \quad "(|\lo|,\ldots, |\lN|)=(|\mo|,\ldots, |\mN|)  
\quad \mathrm{and } \quad 
\lambda^{(i)} \geq \mu^{(i)} \quad (\forall i)" .
\end{align}
Then we have $L^{(n,-)}_{\vl,\vm}=0$ 
unless $\vl \overset{**}{<}^{\mathrm{R}} \vm$.
}
and its diagonal elements are 
\begin{equation}
L_{\vl, \vl}=u_1^{\sum_{i=1}^{N} \ell(\lambda^{(i)})} u_2^{\sum_{i=2}^{N}\ell(\lambda^{(i)})} \cdots u_N^{\ell(\lambda^{(N)}) }, 
\end{equation}
we have 
\begin{align}
\det(L^{(n,-)}_{\vl,\vm})
&=u_1^{\sum_{\vl}\sum_{i=1}^{N} \ell(\lambda^{(i)})} u_2^{\sum_{\vl}\sum_{i=2}^{N}\ell(\lambda^{(i)})} \cdots u_N^{\sum_{\vl}\ell(\lambda^{(N)}) } \\
&=\prod_{k=1}^N u_k^{k \sum_{\vl}\ell(\lambda^{(N)})} \\
&=\prod_{k=1}^N \prod_{\substack{1\leq r,s\\ rs \leq n}} u_k^{k P^{(N)}(n-rs)}. 
\end{align}
The transition matrix $C^{(-)}_{\vl,\vm}$ is upper triangular 
with respect to the partial ordering $\overset{**}{>}^{\mathrm{L}}$, 
and all diagonal elements are $1$. 
Hence, $\det (C^{(-)}_{\vl,\vm})_{\vl,\vm \vdash n}=1$. 
Similarly, with the help of the base transformation to 
$\Bra{\vu} B_{\vl}$, 
it can be seen that 
\begin{equation}
\lc (H^{(n,+)}; u_1, \ldots , u_N)
=\prod_{k=1}^N \prod_{\substack{1\leq r,s\\ rs \leq n}} u_k^{k P^{(N)}(n-rs)}. 
\end{equation}
Therefore the prefactor $g_{N,n}(q,t)$ is  
\begin{equation}
g_{N,n}(q,t)=G_n(q,t)=\prod_{\vl \vdash n} \prod_{k=1}^N b_{\lambda^{(k)}}(q) b'_{\lambda^{(k)}}(t^{-1}). 
\end{equation}
This completes the proof of Theorem \ref{thm:KacDet}. 
%%%%%%%%%%%%%%%%%%%%%%%%%%%%%%%%%%%%%%%%%%%%%%%%%%%%%%%%%%%%%%%%%%
%%%%%%%%%%%%%%%%%%%%%%%%%%%%%%%%%%%%%%%%%%%%%%%%%%%%%%%%%%%%%%%%%%
%%%%%%%%%%%%%%%%%%%%%%%%%%%%%%%%%%%%%%%%%%%%%%%%%%%%%%%%%%%%%%%%%%

%%%%%%%%%%%%%%%%%%%%%%%%%%%%%%%%%%%%%%%%%%%%%%%%%%%%%%%%%%%%%%%%%%
%%%%%%%%%%%%%%%%%%%%%%%%%%%%%%%%%%%%%%%%%%%%%%%%%%%%%%%%%%%%%%%%%%
%%%%%%%%%%%%%%%%%%%%%%%%%%%%%%%%%%%%%%%%%%%%%%%%%%%%%%%%%%%%%%%%%%
\section{Singular vectors and generalized Macdonald functions}
\label{sec:sing vct and Gn Mac}

In this subsection, 
the singular vectors of the algebra $\mathcal{A}(N)$ are discussed. 
Trivially, 
when $u_i=0$, the Kac determinant (\ref{eq:KacDet for DIM}) 
degenerates, 
and it can be easily seen that 
the vectors $a^{(i)}_{-\lambda}\Ket{\vu}$ are singular vectors. 
Since the screening operator $S^{(i)}(z)$ is 
the same one of the deformed $W_N$-algebra, 
the situation of the singular vectors of $\mathcal{A}(N)$ 
except contribution arising when $u_i=0$ 
is the same as the deformed $W_N$-algebra. 
We discover that singular vectors 
obtained by the screening currents $S^{(i)}(z)$ 
coincides with generalized Macdonald functions 
which is also called the AFLT basis.% 
\footnote{
The AFLT basis in 4D AGT conjecture can be constructed by 
the spherical double affine Hecke algebra with central charges 
($SH^c$ algebra). 
The relation between singular vectors of the $SH^{c}$ algebra 
and the AFLT basis is investigated in \cite{FNMZ:2015:SHc}.}

Firstly, 
let us introduce the generalized Macdonald functions. 
To state their existence theorem, 
we prepare the following ordering.

\begin{definition}\label{def:ordering1}
For $N$-tuple of partitions $\vl$ and $\vm$, 
\begin{align}
\vl \overstar{>} \vm \quad \overset{\mathrm{def}}{\Longleftrightarrow} \quad  
& |\vl| = |\vm|, \quad 
\sum_{i=k}^N |\lambda^{(i)}| \geq \sum_{i=k}^N |\mu^{(i)}| \quad (\forall k ) \quad \mathrm{and}   \\
&  (|\lo|,|\lt|,\ldots ,|\lN|) \neq (|\mo|,|\mt|,\ldots ,|\mN|).  \nonumber
\end{align}
Note that the second condition can be replaced 
with 
\begin{equation}
\sum_{i=1}^{k-1} |\lambda^{(i)}| \leq \sum_{i=1}^{k-1} |\mu^{(i)}|\quad  (\forall k ).
\end{equation} 
\end{definition}

The generalized Macdonald functions
are eigenfunctions of the operator $\Xo_0$. 
The operator $\Xo_0$ is triangulated 
in the basis of the product of the ordinary Macdonald functions 
$\prod_{i=1}^N P_{\lambda^{(i)}}(a^{(i)}_{-n};q,t) \Ket{\vu}$ 
arranged in the above ordering, 
where $P_{\lambda}(a_{-n}^{(i)};q,t)$ are Macdonald symmetric functions 
defined in Appendix \ref{sec: Macdonald and HL} 
with substituting the bosons $a^{(i)}_{-n}$ 
for the power sum symmetric functions $p_n$. 
By the triangulation of $\Xo_0$, 
the following existence theorem of the generalized Macdonald functions 
holds.

\begin{fact}[\cite{Ohkubo:2015:Crystallization}]
\label{fact:existence thm of Gn Mac}
For each $N$-tuple of partitions $\vl$, 
there exists a unique vector $\Ket{P_{\vl}} \in \mathcal{F}_{\vu}$ 
such that 
\begin{align}
 &\Ket{P_{\vec{\lambda}}} 
  = \prod_{i=1}^N P_{\lambda^{(i)}}(a^{(i)}_{-n};q,t) \Ket{\vu} 
  + \sum_{\vm \mathop{\overset{*}{<}} \vl} c_{\vl, \vm} \prod_{i=1}^N P_{\mu^{(i)}}(a^{(i)}_{-n};q,t) \Ket{\vu}, \\
 &X^{(1)}_0 \Ket{P_{\vl}} = \epsilon_{\vl} \Ket{P_{\vl}}, 
\end{align}
where $c_{\vl, \vm}=c_{\vl, \vm}(u_1, \ldots ,u_N;q,t)$ is a constant 
and  
$\epsilon_{\vl}=\epsilon_{\vl}(u_1,\ldots,u_N;q,t)$ is the eigenvalue of $X^{(1)}_0$. 
%Similarly, there exists a unique vector 
%$\Bra{P_{\vl}} \in \mathcal{F}_{\vu}^*$ such that 
%\begin{align}
% &\Bra{P_{\vec{\lambda}}} 
%  = \Bra{\vu} \prod_{i=1}^N P_{\lambda^{(i)}}(a^{(i)}_{n};q,t) 
%  + \sum_{\vec{\mu} \overstar{>} \vec{\lambda}} c_{\vl, \vm}^* \Bra{\vu} \prod_{i=1}^N P_{\mu^{(i)}}(a^{(i)}_{n};q,t), \\
% & \Bra{P_{\vec{\lambda}}} \Xo_0 = \epsilon_{\vec{\lambda}}^* \Bra{P_{\vl}}.
%\end{align}
Then the eigenvalues are 
\begin{equation}\label{eq:eigenvalue of gn Mac}
\epsilon_{\vl}= 
%\epsilon_{\vl}^*= 
\sum_{k=1}^N u_k e_{\lambda^{(k)}}, \quad 
e_{\lambda}:= 1+(t-1) \sum_{i \leq 1} (q^{\lambda_i}-1)t^{-i}. 
\end{equation}
\end{fact}

The unique vectors $\Ket{P_{\vl}}$ are called the generalized Macdonald functions or the q-deformed version of the AFLT basis. 
The vectors $\Ket{P_{\vl}}$ is first introduced in \cite{awata2011notes} 
and play an important role in the 5D AGT coorespondence.
Although the ordering of Definition \ref{def:ordering1} 
is different from the one in \cite{awata2011notes}, 
the eigenfunctions $\Ket{P_{\vl}}$ are quite the same. 
By this proposition, 
it can be seen that $\Ket{P_{\vl}}$ is a basis over $\mathcal{F}_{\vu}$, 
and the eigenvalues of $X^{(1)}_0$ are non-degenerate when $u_i$ is generic.

In this paper, 
we discover the relation between 
the vectors $\Ket{P_{\vl}}$ and the singular vectors. 
First, we have the following simple theorem. 
Hereafter, we consider the $N\geq 2$ case.

\begin{theorem}\label{thm:Sing vct and Gn Mac 1}
For a number $i\in \{1, \ldots ,N-1 \}$, 
if $u_{i} = q^st^{-r} u_{i+1}$ and the other $u_j$ are generic, 
there exists a unique singular vector 
$\ket{\chi^{(i)}_{r,s}}$ in $\mathcal{F}_{\vu}$ up to scalar multiple, 
and it corresponds to the generalized Macdonald function $\Ket{P_{\vl}}$ 
with 
\begin{equation}
\vl= (\overbrace{\emptyset , \ldots , \emptyset , (s^r)}^{i+1}, \emptyset, \ldots , \emptyset).
\end{equation}
That is, 
\begin{eqnarray}
\ket{\chi^{(i)}_{r,s}} \propto \Ket{P_{(\emptyset , \ldots , \emptyset , (s^r), \emptyset, \ldots , \emptyset)}}. 
\end{eqnarray}
\end{theorem}

\proof
Existence and uniqueness are understood by the formula 
for the Kac determinant (\ref{eq:KacDet for DIM}) in the usual way. 
In fact, 
the unique singular vector $\ket{\chi^{(i)}_{r,s}}$ is 
the one of (\ref{eq:sing vct chi(i)rs}). 
Since the screening charges commute with $X^{(1)}_0$, 
the singular vector is an eigenfunction of $\Xo_0$ of the eigenvalue 
$\sum_{i=1}^N v_i$. 
Using the relations 
$u_k=v_k$ for $k \neq i, i+1$ and 
$u_{i+1}=t^r v_{i+1}$, $u_{i}=t^{-r} v_{i}$, 
we have 
\begin{equation}
\sum_{i=1}^N v_i 
= \epsilon_{(\emptyset , \ldots , \emptyset , (s^r), \emptyset, \ldots , \emptyset)}(u_1, \ldots , u_N), 
\end{equation}
where $\epsilon_{\vl}=\epsilon_{\vl}(u_1, \ldots , u_N)$ is the eigenvalue
of the generalized Macdonald functions 
introduced in (\ref{eq:eigenvalue of gn Mac}). 
Thus, 
the singular vector $\ket{\chi^{(i)}_{r,s}}$ 
and the generalized Macdonald function 
$\Ket{P_{(\emptyset , \ldots , \emptyset , (s^r), \emptyset, \ldots , \emptyset)}}$ 
are in the same eigenspace of $\Xo_{0}$. 
Moreover, 
by comparing the eigenvalues $\epsilon_{\vl}$, 
it can be shown that the dimension of the eigenspace of the eigenvalue 
$\epsilon_{(\emptyset , \ldots , \emptyset , (s^r), \emptyset, \ldots , \emptyset)}$ 
is $1$ even when $u_{i} = q^st^{-r} u_{i+1}$. 
Therefore, 
this theorem follows. 
\qed

Let us consider more complicated cases. 
For variables $\alpha^{(k)}$ ($k=1,\ldots, N-1$), 
define the function $\check{h}^{(i)}$ by 
\begin{align}
\check{h}^{(i)}(\alpha^{(k)}) \seteq 
\frac{1}{N} &\left( \alpha^{(1)}+2\alpha^{(2)}+ \cdots +(i-1) \alpha^{(i-1)} \right. \nonumber \\
&\left. -(N-i)\alpha^{(i)} -(N-i-1)\alpha^{(i+1)} -\cdots - \alpha^{(N-1)} \right). 
\end{align}
Then it satisfies 
$\alpha^{(i)}=\check{h}^{(i+1)}(\alpha^{(k)})-\check{h}^{(i)}(\alpha^{(k)})$. 
We focus on the following singular vectors 
\begin{align}
\Ket{\chi_{\vec{r}, \vec{s}}} := 
\oint \prod_{k=1}^{N-1} \prod_{i=1}^{r_k} dz^{(k)}_i 
&S^{(N-1)}(z^{(N-1)}_1) \cdots S^{(N-1)}(z^{(N-1)}_{r_{N-1}}) \cdots \nonumber \\
& \cdots S^{(1)}(z^{(1)}_1) \cdots S^{(1)}(z^{(1)}_{r_{1}}) 
\ket{\vv}, 
\end{align}
where 
the parameter $\vv=(v_1,\ldots,v_N)$ is 
$v_i=v'' v'_i$, 
$v'_i=q^{\sqrt{\beta}\check{h}^{(i)}(\widetilde{\alpha}^{k}_{\vec{r}, \vec{s}}) } p^{-\frac{N+1}{2}+i}$, and 
for non-negative integers $s_k>0$ and $r_k$ ($k=1,\ldots, N-1$), 
\begin{equation}
\tilde{\alpha}^{(k)}_{\vr, \vs}\seteq 
\sqrt{\beta} (1-r_k +r_{k-1})-\frac{1}{\sqrt{\beta}}(1+s_k),\quad 
r_0:=0. 
\end{equation}
The singular vector $\Ket{\chi_{\vec{r},\vec{s}}}$ 
is in the Fock module $\mathcal{F}_{\vu}$ 
of the highest weight $\vu=(u_1,\ldots,u_N)$ defined by $u_i=u'' u'_i$, 
$u'_i=q^{\sqrt{\beta}\check{h}^{(i)}(\alpha^{k}_{\vec{r}, \vec{s}}) } p^{-\frac{N+1}{2}+i}$, $u''=v''$ and 
\begin{equation}\label{eq:alpha for sing. vct}
\alpha^{(k)}_{\vec{r},\vec{s}} \seteq 
\sqrt{\beta} (1+r_k -r_{k+1})-\frac{1}{\sqrt{\beta}}(1+s_k),\quad 
r_N:=0.
\end{equation} 
The equation 
(\ref{eq:alpha for sing. vct}) is equivalent to the condition 
$u_i=q^{s_i}t^{-r_i+r_{i+1}}u_{i+1}$ ($\forall i$) of parameters $u_i$. 
We can also see that 
$\ket{\vu}=e^{\sum_{i=1}^{N-1}\alpha_{\vr,\vs}^{(i)} \Qi_{\Lambda}
+\alpha' Q'} \ket{0}$ 
under the parametrization $u''=q^{\sqrt{\beta}\frac{\alpha'}{N}}$. 
Here, $\Qi_{\Lambda}$ and $Q'$ are defined in Section \ref{sec:Proof of Kacdet}.% 
\footnote{
The screening currents $S^{(i)}(z)$, 
the parameters $\alpha^{(k)}_{\vr,\vs}$ and integers $r_i$, $s_i$ in this paper 
correspond to $S^{N-i}_+(z)$, $\alpha^{N-k}_{r,s}$, $r_{N-i}$ and $s_{N-i}$ in \cite{Awata:1995Quantum}, respectively.  
}

To introduce the partition of a generalized Macdonald function 
corresponding to this singular vector, 
we prepare following notations. 
For a partition $\lambda=(\lambda_1,\lambda_2,\ldots)$ and a non-negative 
integer $n <\ell(\lambda)$, 
we define
\begin{equation}
\mathsf{T}(\lambda; n):=(\lambda_1,\lambda_2,\ldots, \lambda_n), \quad 
\mathsf{R}(\lambda; n):=(\lambda_{n+1}, \lambda_{n+2},\ldots,\lambda_{\ell(\lambda)}). 
\end{equation}
When $n\geq \ell(\lambda)$, we put 
$\mathsf{T}(\lambda;s):=\lambda$, $\mathsf{R}(\lambda;s):=\emptyset$. 
For partitions $\lambda$ and $\mu$ satisfying 
$\lambda_{\ell(\lambda)} \geq \mu_1$, define
\begin{equation}
\mathsf{J}(\lambda,\mu):=(\lambda_1, \lambda_2,\ldots , \lambda_{\ell(\lambda)},\mu_1,\mu_2,\ldots ). 
\end{equation}
In this paper, 
for an $n$-tuple of integers $\vec{k}=(k_1, \ldots , k_n)$, 
let $\vec{k}^{*}$ %with the star 
denote the $(n-1)$-tuple of integers except the last integer, 
i.e., 
$\vec{k}^{*}:=(k^{*}_1, \ldots, k^{*}_{n-1}):=(k_1, \ldots, k_{n-1})$, 
and we put $k^{*}_n=0$.

For $(N-1)$-tuples of non-negative integers 
$\vr=(r_{1},\ldots,r_{N-1})$ and $\vs=(s_{1},\ldots,s_{N-1})$, 
the $N$-tuple of partitions 
$\Theta_{\vr,\vs}=(\theta_{\vr,\vs}^{(1)},\ldots,\theta^{(N)}_{\vr,\vs})$ 
is inductively defined by 
$\Theta_{(r_1),(s_1)}:=(\emptyset,(s_1^{r_1}))$ and 
\begin{equation}
\Theta_{\vr,\vs}:=\left(\theta^{(1)}_{\vr^{*},\vs^{*}},\ldots, 
\theta^{(N-2)}_{\vr^{*},\vs^{*}},\mathsf{R}(\theta^{(N-1)}_{\vr^{*},\vs^{*}};r_{N-1}),
\mathsf{T}(\theta^{(N-1)}_{\vr^{*},\vs^{*}};r_{N-1})+((s_{N-1})^{r_{N-1}})\right). 
\end{equation}
The meaning of this $N$-tuple of partitions will be described 
in Example \ref{ex:Gn Mac and sing. vct 1}.

We obtain the following theorem which states 
the coincidence of the singular vectors and 
generalized Macdonald functions with the $N$-tuple of partitions $\Theta_{\vr,\vs}$. 
This theorem is an extension of the result in \cite{Awata:1995Quantum}.

\begin{theorem}\label{thm:Sing vct and Gn Mac 2}
Let $\vs$ be an $(N-1)$-tuple of positive integers and 
$\vr$ be an $(N-1)$-tuple of non-negative integers. 
If parameters $u_i$ satisfy 
$u_{i} = q^{s_{i}} t^{-r_{i}+r_{i+1}} u_{i+1}$ for all $i$ 
($r_{N}:=0$), 
then 
the singular vector $\Ket{\chi_{\vec{r},\vec{s}}}$ 
coincides with the generalized Macdonald function 
$\Ket{P_{\Theta_{\vr,\vs}}}$, i.e., 
\begin{equation}
\Ket{\chi_{\vec{r},\vec{s}}} \propto 
\Ket{P_{\Theta_{\vr,\vs}}}. 
\end{equation}
\end{theorem}

Before the proof of this theorem, 
we note that the eigenvalues $e_{\lambda}$ of ordinary Macdonald functions
can be written by contribution from the edges of the Young diagram:
\begin{equation}\label{eq:edge contribution of e_lambda}
e_{\lambda}=
\sum_{(i,j)\in A(\lambda)}q^{j-1}t^{-i+1}-\sum_{(i,j)\in R(\lambda)} q^{j}t^{-i},
\end{equation}
where $A(\lambda)$ and $R(\lambda)$ are 
the sets of coordinates of the boxes 
which can be added to or removed 
from Young diagram of $\lambda$, respectively. 
We prepare a simple Lemma with respect to the eigenvalues $e_{\lambda}$.

\begin{lemma}\label{lem:properites of eigenvalue}
For a partition $\lambda$ with $\ell(\lambda)\leq r$
\begin{equation}
e_{\lambda+(s^r)}=q^s e_{\lambda}-q^st^{-r}+t^{-r}.
\end{equation}
For a partition $\lambda$ and a non-negative integer $n$, 
\begin{equation}\label{eq:lem of e_lambda 2}
e_{\lambda}=
e_{\mathsf{T}(\lambda;n)}+
t^{-r}e_{\mathsf{R}(\lambda;n)}-t^{-n}.
\end{equation}
Using partitions $\lambda$ and $\mu$ 
satisfying $\lambda_{\ell(\lambda)}\leq \mu_1$, 
the equation (\ref{eq:lem of e_lambda 2}) can be also written as 
\begin{equation}
e_{\mathsf{J}(\lambda,\mu)}=
e_{\lambda}+
t^{-\ell(\lambda)}e_{\mu}-t^{-\ell(\lambda)}. 
\end{equation}
\end{lemma}

This lemma is easily proved by the expression 
(\ref{eq:edge contribution of e_lambda}) of 
eigenvalues $e_{\lambda}$. 
We now prove Theorem \ref{thm:Sing vct and Gn Mac 2}.

\noindent 
\textit{Proof of Theorem \ref{thm:Sing vct and Gn Mac 2}.}
The method of the proof is similar 
to Theorem \ref{thm:Sing vct and Gn Mac 1}. 
Using the fact that generalized Macdonald functions and 
the singular vector $\Ket{\chi_{\vr,\vs}}$ are 
eigenfunctions of the operator $\Xo_0$, 
we can prove this Theorem by comparison of each eigenvalue.

First, we prove that the eigenvalue $\epsilon_{\Theta_{\vr,\vs}}$ 
of the generalized Macdonald function coincides with 
the eigenvalue $\sum_{i=1}^N v_i$ 
of the singular vector $\Ket{\chi_{\vr,\vs}}$. 
It is useful to show the coincidence of each quotient by $u_N$, i.e.,  
\begin{equation}\label{eq:eigenvalue of Gn Mac/u_N}
\frac{1}{u_N} \epsilon_{\Theta_{\vr,\vs}}
=\sum_{i=1}^N\frac{u_i}{u_N}e_{\theta^{(i)}_{\vr,\vs}}
=\sum_{i=1}^N t^{-r_i}q^{s_i+\cdots + s_{N-1}} e_{\theta^{(i)}_{\vr,\vs}}
\end{equation}
and
\begin{equation}\label{eq:eigenvalue of Sing vct/u_N}
\frac{1}{u_N}\sum_{i=1}^N v_i
=\sum_{i=1}^N t^{-r_{i-1}}q^{s_i+\cdots +s_{N-1}}. 
\end{equation}
Here, $s_i+ \cdots +s_{N-1}=0$ for $i=N$. 
We use the mathematical induction on $N$. 
If $N=2$, it is clear that 
\begin{equation}
\frac{1}{u_2} \epsilon_{\Theta_{(r_1),(r_2)}}
=t^{-r_1}q^{s_1} e_{\emptyset} + e_{(s_1^{r_1})}
= q^{s_1} +t^{-r_1}
=\sum_{i=1}^2 \frac{v_i}{u_2}, 
\end{equation}
Assuming the coincidence of (\ref{eq:eigenvalue of Gn Mac/u_N}) 
and (\ref{eq:eigenvalue of Sing vct/u_N}) for $N=n-1$, 
we will prove it for $N=n$. 
By definition of $\Theta_{\vr,\vs}$ and Lemma \ref{lem:properites of eigenvalue}, 
\begin{align}
\frac{1}{u_N}\epsilon_{\Theta_{\vr,\vs}} 
&=t^{-r_{N-1}}q^{s_{N-1}}
\left\{ \sum_{i=1}^{N-2}
\frac{u_i}{u_{N-1}}e_{\theta^{(i)}_{\vr^{*},\vs^{*}}}
+e_{\mathsf{R}(\theta^{(N-1)}_{\vr^{*},\vs^{*}};r_{N-1})} \right. \nonumber \\
&\qquad \left. +\frac{u_{N}}{u_{N-1}}e_{\mathsf{T}(\theta^{(N-1)}_{\vr^{*},\vs^{*}};r_{N-1})+(s_{N-1}^{r_{N-1}})}
\right\} \nonumber \\
&=q^{s_{N-1}}
\left\{ \sum_{i=1}^{N-2}
t^{-r^{*}_i}q^{s^{*}_i+\cdots+s^{*}_{n-2}} e_{\theta^{(i)}_{\vr^{*},\vs^{*}}}
+t^{-r_{N-1}} e_{\mathsf{R}(\theta^{(N-1)}_{\vr^{*},\vs^{*}};r_{N-1})}\right. \nonumber \\
& \qquad +e_{\mathsf{T}(\theta^{(N-1)}_{\vr^{*},\vs^{*}};r_{N-1})} 
-t^{-r_{N-1}}+q^{-s_{N-1}}t^{-r_{N-1}}
\biggr\} \nonumber \\
&=q^{s_{N-1}}
\left\{ \sum_{i=1}^{N-2}
t^{-r^{*}_i}q^{s^{*}_i+\cdots+s^{*}_{n-2}} e_{\theta^{(i)}_{\vr^{*},\vs^{*}}}
+e_{\theta^{(N-1)}_{\vr^{*},\vs^{*}}}
+q^{-s_{N-1}}t^{-r_{N-1}}
\right\} \nonumber \\
&=q^{s_{N-1}}
\left\{ \sum_{i=1}^{N-1}
t^{-r^{*}_i}q^{s^{*}_i+\cdots+s^{*}_{n-2}} e_{\theta^{(i)}_{\vr^{*},\vs^{*}}} 
\right\}
+t^{-r_{N-1}}.  \label{eq:cal. of eigenvalue}
\end{align}
By assumption for $N=n-1$, 
we can see that (\ref{eq:cal. of eigenvalue}) is 
\begin{equation}
q^{s_{N-1}}
\left\{ \sum_{i=1}^{N-1}
t^{-r_{i-1}}q^{s_i+\cdots+s_{n-2}}  
\right\}
+t^{-r_{N-1}}
=\sum_{i=1}^{N}t^{N-1}q^{s_{N-i+1}+\cdots+s_{N-1}}
=\sum_{i=1}^{N}\frac{v_i}{u_N}. 
\end{equation}
Therefore, by induction on $N$, 
the eigenvalue of the generalized Macdonald function 
$\Ket{P_{\Theta_{\vr,\vs}}}$
coincides with the one of the singular vector $\Ket{\chi_{\vr,\vs}}$ 
for general $N$. 
By this fact, 
$\Ket{P_{\Theta_{\vr,\vs}}}$ and $\Ket{\chi_{\vr,\vs}}$ 
belong to the same eigenspace of $\Xo_0$.

The proof of this Theorem is completed by showing that 
the dimension of the eigenspace of the eigenvalue $\sum_{i=1}^N v_i$ 
is $1$. 
Since we know all eigenvalues of $\Xo_0$ 
by Fact \ref{fact:existence thm of Gn Mac}, 
it is sufficient to prove that 
if an $N$-tuple of partitions $\vl=(\lo,\ldots,\lN)$
satisfies $\epsilon_{\vl}=\sum_{i=1}^N v_i$, 
then $\vl=\Theta_{\vr,\vs}$. 
That is to say, we will show 
\begin{equation}\label{eq:dim of eigensp is 1}
\epsilon_{\vl}=\sum_{i=1}^N v_i \quad 
\Longrightarrow \quad 
\vl=\Theta_{\vr,\vs}. 
\end{equation}
Actually, 
we consider the quotient by $u_N$ instead of the equation 
$\epsilon_{\vl}=\sum_{i=1}^N v_i$ in (\ref{eq:dim of eigensp is 1}):
\begin{equation}\label{eq:e/u_N=v /u_N}
%\frac{1}{u_N}\epsilon_{\lambda}
\sum_{i=1}^N t^{-r_i}q^{s_i+\cdots + s_{N-1}} e_{\lambda^{(i)}}
=\sum_{i=1}^{N} t^{-r_{i-1}}q^{s_i+\cdots+s_{N-1}}. 
\end{equation}
Let us denote by $E_i$ each term of LHS in (\ref{eq:e/u_N=v /u_N}), 
i.e., 
\begin{equation}
E_i :=t^{-r_i}q^{s_i+\cdots + s_{N-1}} e_{\lambda^{(i)}}. 
\end{equation}

First, we deal with the case that $r_{N-1}\neq 0$. 
Let $\vl$ satisfy the equation (\ref{eq:e/u_N=v /u_N}) 
and $i_0 \leq N-1$ 
be the maximum number such that $r_{N-1}> r_{i_0}$  
($r_0:=0$). 
Then we have 
\begin{equation}\label{eq:cond. of lN}
\lN_{r_{N-1}}=s_{i_0+1}+\cdots +s_{N-1} \quad 
\mathrm{and} \quad 
\ell(\lN) =r_{N-1}.
\end{equation}
The property (\ref{eq:cond. of lN}) is shown as follows. 
We focus on the term 
\begin{equation}
t^{-r_{i_0}}q^{s_{i_0+1}+\cdots + s_{N-1}} =: \mathcal{T}_{i_0}
\end{equation}
in the RHS of (\ref{eq:e/u_N=v /u_N}). 
In order that the LHS reproduces the term $\mathcal{T}_{i_0}$, 
it is necessary that there is a number $j_0$ such that 
the expansion of $E_{j_0}$ contains the term $\mathcal{T}_{i_0}$. 
We can see $j_0=N$, 
because for a number $j$ such that $i_0 \geq j$, 
the degree with respect to $q$ of $E_j$ 
is larger than the degree $s_{i_0+1}+\cdots + s_{N-1}$ 
of $\mathcal{T}_{i_0}$
and for a number $j$ such that $N-1 \geq j > i_0$, 
the degree with respect to $t$ of $E_j$ 
is smaller than $-r_{i_0}$ 
owing to the maximality of $i_0$. 
Therefore, 
the only term $E_N=e_{\lN}$ in the LHS of (\ref{eq:e/u_N=v /u_N}) 
contains $\mathcal{T}_{i_0}$. 
Since $e_{\lambda}$ can be written by contribution from 
the edges of Young diagram as (\ref{eq:edge contribution of e_lambda}), 
the partition $\lN$ satisfies 
\begin{equation}\label{eq:condition of lN 1}
A(\lN) \ni (r_{i_0}+1,s_{i_0+1}+\cdots +s_{N-1}+1). 
\end{equation}
By (\ref{eq:condition of lN 1}), 
\begin{equation}\label{eq:condition of lN 2}
R(\lN) \ni (r_{i_0}+m_{\check{s}},s_{i_0+1}+\cdots +s_{N-1}), 
\end{equation}
where $\check{s} := s_{i_0+1}+\cdots +s_{N-1}$ and 
$m_{\check{s}}=m_{\check{s}}(\lN)$ is the number of entries in $\lN$ 
that are equal to $\check{s}$. 
Hence, the term 
\begin{equation}\label{eq:minus term}
-t^{-(r_{i_0}+m_{\check{s}})}q^{s_{i_0+1}+\cdots +s_{N-1}}
\end{equation}
is contained in $e_{\lN}$. 
Since there is no term with the minus sign
in the RHS of (\ref{eq:e/u_N=v /u_N}), 
$m_{\check{s}}$ should be a value such that 
the term (\ref{eq:minus term})
is canceled by other terms arising from 
$E_i$ ($i\neq N$). 
The degree with respect to $q$ of $E_i$ 
for $i_0 \geq i$ is larger than the one of (\ref{eq:minus term}). 
In addition, 
the maximum degree with respect to $t$ of $E_i$ ($N>i>i_0$) 
is $\max \{ -r_{i_0+1}, \ldots , -r_{N-1}\} = -r_{N-1}$ 
by the maximality of $i_0$. 
Therefore, we have 
\begin{equation}\label{eq:size of m_checks}
r_{i_0} + m_{\check{s}}(\lN)\geq r_{N-1}.   
\end{equation}
From (\ref{eq:condition of lN 2}) and (\ref{eq:size of m_checks}),
it can be seen that $\lN_{r_{N-1}}=s_{i_0+1}+\cdots s_{N-1}$. 
On the other hand, 
in order to reproduce the term $t^{-r_{N-1}}$ in the RHS 
of (\ref{eq:e/u_N=v /u_N}), 
it is necessary that $\lN$ satisfy 
\begin{equation}\label{eq:condition of lN 3}
A(\lN) \ni (r_{N-1}+1, 1), 
\end{equation}
because the degree with respect to $q$ 
of $E_i$ for $i\neq N$ 
is too large to reproduce the term $t^{-r_{N-1}}$. 
By (\ref{eq:condition of lN 3}), 
we have $\ell(\lN)=r_{N-1}$ and $r_{i_0} + m_{\check{s}}(\lN)=r_{N-1}$. 
Thus (\ref{eq:cond. of lN}) is proved.

Next, let us show that
\begin{equation}\label{eq:cond. of lN-1}
\lambda^{(N-1)}_1 \leq \lN_{r_{N-1}}-s_{N-1}. 
\end{equation}
(\ref{eq:cond. of lN-1}) is proved as follows. 
When there exists a number $i$ such that 
\begin{equation}\label{eq:cond. of i_e}
r_{N-1}=r_{i} \quad \mbox{and} \quad N-1>i>i_0, 
\end{equation}
let $i_e$ be the maximum number satisfying the condition 
(\ref{eq:cond. of i_e}). 
There is a number $j_e$ such that 
$E_{j_e}$ contains the term 
\begin{equation}
t^{-r_{i_e}}q^{s_{i_e+1}+\cdots+s_{N-1} } =:\mathcal{T}_{i_e}  
\end{equation} 
appearing in RHS of (\ref{eq:e/u_N=v /u_N}). 
Then $j_e$ satisfies
\begin{equation}
j_e \geq i_e +1
\end{equation}
by the comparison of degrees of $q$. 
However, $j_e$ does not satisfy 
\begin{equation}
N-1 >j_e \geq i_e+1
\end{equation}
because the degree of $t$ of $E_j$ 
($N-1 >j \geq i_e+1$) 
is smaller than $-r_{i_e}=-r_{N-1}$ 
owing to the maximality of $i_0$ and $i_e$. 
Furthermore, $j_e \neq N$ by the property (\ref{eq:cond. of lN}). 
Therefore $j_e=N-1$. 
Since the only term $E_{N-1}=t^{r_{N-1}}q^{s_{N-1}} e_{\lambda^{(N-1)}}$
contains $\mathcal{T}_{i_e}$, 
we can see that $\lambda^{(N-1)}_1 = s_{i_e+1}+\cdots +s_{N-2}$. 
Here, 
if $i_e=N-2$, then $\lambda^{(N-1)}_1=0$, i.e., $\lambda^{(N-1)}=\emptyset$. 
Next, when there is no number such that (\ref{eq:cond. of i_e}), 
a number $j$ such that 
$E_j$ 
can cancel the term (\ref{eq:minus term}) with the minus sign  
is only $N-1$ by similar discussion using degrees of $q$ and $t$.  
Hence we get $\lambda^{(N-1)}_1=s_{i_0+1}+\cdots +s_{N-2}$. 
Thus (\ref{eq:cond. of lN-1}) is proved.

By the above properties, 
we are ready to write the following equation: 
\begin{align}\label{eq:cal. of epsilon_vl}
\frac{1}{u_N}\epsilon_{\vl}
&= \frac{1}{u_N}\sum_{i=1}^{N-1} u_i e_{\lambda^{(i)}} 
%+ t^{-r_{N-1}}q^{s_{N-1}} e_{\lambda^{(N-1)}}
+q^{s_{N-1}}e_{\lN-(s_{N-1}^{r_{N-1}})} 
-t^{-r_{N-1}}q^{s_{N-1}}+ t^{-r_{N-1}} \nonumber \\
&= \frac{1}{u_N}\sum_{i=1}^{N-2} u_i e_{\lambda^{(i)}} 
+q^{s_{N-1}}e_{\mathsf{J}(\lN-(s_{N-1}^{r_{N-1}}),\lambda^{(N-1)})} 
+ t^{-r_{N-1}}, 
\end{align}
where 
$\mathsf{J}(\lN-(s_{N-1}^{r_{N-1}}),\lambda^{(N-1)})$ is 
well-defined by the properties 
(\ref{eq:cond. of lN}) and (\ref{eq:cond. of lN-1}). 
Put $\mu^{(i)}:=\lambda^{(i)}$ ($i=1,\ldots, N-2$), 
$\mu^{(N-1)}:=\mathsf{J}(\lN-(s_{N-1}^{r_{N-1}}),\lambda^{(N-1)})$. 
Combining (\ref{eq:cal. of epsilon_vl}) with (\ref{eq:e/u_N=v /u_N}) 
yields  
\begin{equation}\label{eq:N-1 case}
\sum_{i=1}^{N-1} t^{-r_i} q^{s_i+\cdots + s_{N-2}}e_{\mu^{(i)}}
=\sum_{i=1}^{N-1} t^{-r_{i-1}}q^{s_{i}+\cdots+s_{N-2}}. 
\end{equation}
We dealt with the case that $r_{N-1}\neq 0$. 
Also in the case that $r_{N-1}=0$, 
it is clear that $\lambda^{(N)}=\emptyset$,  
and we can get the same equation as (\ref{eq:N-1 case}). 
Assuming (\ref{eq:dim of eigensp is 1}) holds for the case obtained by replacing $N$ with $N-1$, 
we obtain 
\begin{equation}
\vm=\Theta_{\vr^{*},\vs^{*}}. 
\end{equation}
Hence, $\lambda^{(i)}=\theta^{(i)}_{\vr^{*},\vs^{*}}$ for $i\leq N-2$,  
and 
\begin{equation}
\mathsf{J}(\lN-(s_{N-1}^{r_{N-1}}),\lambda^{(N-1)})=\theta^{(N-1)}_{\vr^{*},\vs^{*}}. 
\end{equation} 
Since $\ell(\lN)=r_{N-1}$, 
\begin{equation}
\lambda^{(N-1)}=\mathsf{R}(\theta^{(N-1)}_{\vr^{*},\vs^{*}};r_{N-1}), \quad 
\lN-(s_{N-1}^{r_{N-1}})
=\mathsf{T}(\theta^{(N-1)}_{\vr^{*},\vs^{*}};r_{N-1}). 
\end{equation}
Therefore, we have $\vl=\Theta_{\vr,\vs}$. 
By induction on $N$, 
(\ref{eq:dim of eigensp is 1}) holds for general $N$. 
This completes the proof. 
\qed

\begin{figure}[H]
\begin{center}
\includegraphics[width=12cm]{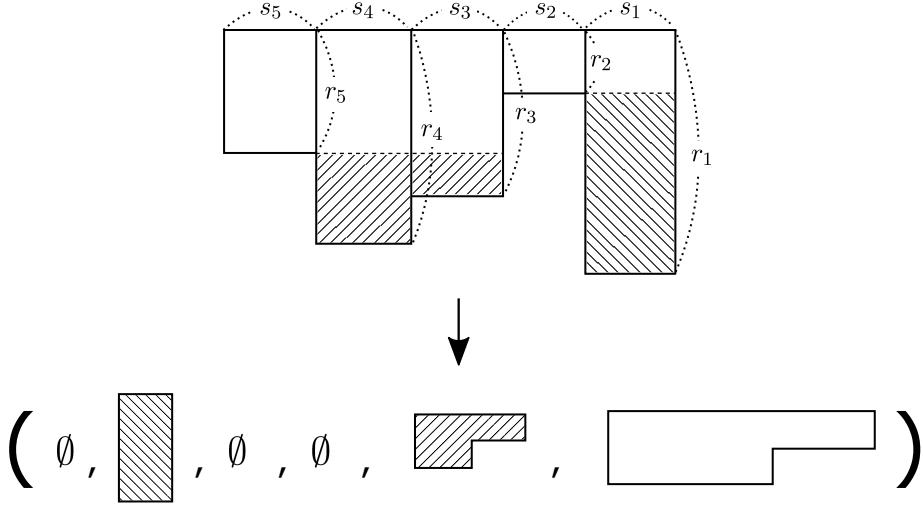}
\caption{The example of $\Theta_{\vr,\vs}$ when $N=6$ and $r_1>r_4>r_3>r_5>r_2$. }
\label{fig:Young2}
\end{center}
\end{figure}

\begin{example}\label{ex:Gn Mac and sing. vct 1}
If $r_{k+1} \geq r_{k} \geq 0$ for all $k$, then 
the singular vector $\Ket{\chi_{\vec{r},\vec{s}}}$ 
coincides with the generalized Macdonald function 
$\Ket{P_{(\emptyset,\ldots ,\emptyset, \lrs)}}$
with
\begin{equation} 
\lrs:= ((s_1+\cdots +s_{N-1})^{r_{1}}, (s_2+\cdots+s_{N-2})^{r_{2}-r_{1}},\ldots, s_{N-1}^{r_{N-1}-r_{N-2}} ). 
\end{equation}
$\lrs$ corresponds to the Young diagram with $N-1$ edges 
in Figure \ref{fig:YoungDiag_OnlyRightSide} in Introduction. 
If $r_k$ does not satisfy the condition 
$r_{k+1} \geq r_{k} \geq 0$, 
Figure \ref{fig:YoungDiag_OnlyRightSide} is not a Young diagram. 
In this case, 
the Young diagram corresponding to the singular vector 
$\Ket{\chi_{\vr,\vs}}$
is obtained by cutting off the protruding part and 
moving the boxes to Young diagrams on the left side. 
This is the meaning of the $N$-tuple of partitions $\Theta_{\vr,\vs}$. 
For example, 
when $N=6$ and $r_1>r_4>r_3>r_5>r_2$, 
the $N$-tuple of Young diagrams $\Theta_{\vr,\vs}$ 
is the one pictured in Figure \ref{fig:Young2}. 
%\begin{equation}
%(\emptyset, (s_1^{r_1-r_2}),\emptyset, \emptyset, 
%((s_3+s_4)^{r_3-r_5},s_4^{r_4-r_3}), 
%((s_1+\cdots +s_5)^{r_2},(s_3+s_4+s_5)^{r_5-r_2})). 
%\end{equation}
\end{example}

It is known that if $r_{k+1} \geq r_{k} \geq 0$ for all $k$,
the projection of the singular vectors $\Ket{\chi_{\vr,\vs}}$ 
onto the diagonal components of the boson $h^{(N)}_n$ 
corresponds to ordinary Macdonald functions \cite[(35)]{Awata:1995Quantum}.  
Hence, 
ordinary Macdonald functions are obtained 
by the projection of generalized Macdonald functions. 
\begin{corollary}\label{cor:projection of Gn Mac}
When $u_{i} = q^{s_{i}} t^{-r_{i}+r_{i+1}} u_{i+1}$ for all $i$, 
\begin{equation}
P_{\lrs}(p_n;q,t) \propto 
\Bra{\vu} \exp \left\{ -\sum_{n>0}p_n \frac{h^{(N)}_n}{1-q^{n}} \right\} \ket{P_{(\emptyset,\ldots ,\emptyset, \lrs)}}. 
\end{equation}
Here, $p_n$ denotes the ordinary power sum symmetric functions. 
\end{corollary}

%%%%%%%%%%%%%%%%%%%%%%%%%%%%%%%%%%%%%%%%%%%%%%%%%%%%%%%%%%%%%%%%%%
%%%%%%%%%%%%%%%%%%%%%%%%acknowledgments%%%%%%%%%%%%%%%%%%%%%%%%%%%
%%%%%%%%%%%%%%%%%%%%%%%%%%%%%%%%%%%%%%%%%%%%%%%%%%%%%%%%%%%%%%%%%%
\section*{Acknowledgments}
The author show his greatest appreciation to 
H. Awata, M. Bershtein, B. Feigin, P. Gavrylenko, K. Hosomichi, H. Itoyama, H. Kanno, A. Marshakov, T. Matsumoto, A. Mironov, S. Moriyama, Al. Morozov, An. Morozov, H. Nagoya, A. Negut, T. Okazaki, T. Shiromizu, T. Takebe, M. Taki, S. Yanagida and Y. Zenkevich 
for variable discussions and comments. 
The author is supported in part by Canon Foundation Research Fellowship. 
%%%%%%%%%%%%%%%%%%%%%%%%%%%%%%%%%%%%%%%%%%%%%%%%%%%%%%%%%%%%%%%%%%
%%%%%%%%%%%%%%%%%%%%%%%%%%%%%%%%%%%%%%%%%%%%%%%%%%%%%%%%%%%%%%%%%%
%%%%%%%%%%%%%%%%%%%%%%%%%%%%%%%%%%%%%%%%%%%%%%%%%%%%%%%%%%%%%%%%%%

\appendix
%%%%%%%%%%%%%%%%%%%%%%%%%%%%%%%%%%%%%%%%%%%%%%%%%%%%%%%%%%%%%%%%%%
%%%%%%%%%%%%%%%%%%%%%%%%%%Appendix%%%%%%%%%%%%%%%%%%%%%%%%%%%%%%%%
%%%%%%%%%%%%%%%%%%%%%%%%%%%%%%%%%%%%%%%%%%%%%%%%%%%%%%%%%%%%%%%%%%
\section*{Appendix}

\section{Macdonald functions }\label{sec: Macdonald and HL}

In this Appendix, 
we give the definition of the ordinary Macdonald functions \cite[Chap.\ VI]{Macdonald}.

Let $\Lambda$ be the ring of symmetric functions, 
and  
$p_{\lambda} = \prod_{k\geq 1} p_{\lambda_k}$ 
($p_n =\sum_{i\geq 1} x_i^n$) is the power sum symmetric functions. 
Define the inner product $\langle-,- \rangle_{q,t}$ over $\Lambda$ 
by the condition that 
\begin{equation}
\langle p_{\lambda}, p_{\mu} \rangle_{q,t} 
= z_{\lambda} \prod_{k=1}^{\ell(\lambda)} \frac{1-q^{\lambda_k}}{1-t^{\lambda_k}} \delta_{\lambda, \mu}, \quad 
z_{\lambda} \seteq \prod_{i \geq 1} i^{m_i} m_i !,
\end{equation}
where $m_i=m_i (\lambda)$ is the number of entries in $\lambda$ equal $i$. 
For a partition $\lambda$, 
Macdonald functions $P_{\lambda} \in \Lambda$ are defined to be 
the unique functions in $\Lambda$ 
satisfying the following two conditions:
\begin{align}
 &\lambda  \neq \mu \quad \Rightarrow \quad \langle P_{\lambda}, P_{\mu} \rangle_{q,t}  = 0;  \\
 &P_{\lambda} = m_{\lambda} + \sum_{\mu < \lambda} c_{\lambda \mu} m_{\mu}.  
\end{align}
Here $m_{\lambda}$ is a monomial symmetric function and 
$<$ is the ordinary dominance partial ordering.  
In this paper, 
the power sum symmetric functions $p_n$ ($n \in \mathbb{N}$) 
are regarded as the variables of Macdonald functions. 
That is to say, $P_{\lambda}= P_{\lambda}(p_n; q,t)$. 
Here $P_{\lambda}(p_n; q,t)$ is an abbreviation for $P_{\lambda}(p_1, p_2, \ldots; q,t)$. 
%%%%%%%%%%%%%%%%%%%%%%%%%%%%%%%%%%%%%%%%%%%%%%%%%%%%%%%%%%%%%%%%
%%%%%%%%%%%%%%%%%%%%%%%%%%%%%%%%%%%%%%%%%%%%%%%%%%%%%%%%%%%%%%%%
%%%%%%%%%%%%%%%%%%%%%%%%%%%%%%%%%%%%%%%%%%%%%%%%%%%%%%%%%%%%%%%%

%%%%%%%%%%%%%%%%%%%%%%%%%%%%%%%%%%%%%%%%%%%%%%%%%%%%%%%%%%%%%%%%
%%%%%%%%%%%%%%%%%%%%%%%%%%%%%%%%%%%%%%%%%%%%%%%%%%%%%%%%%%%%%%%%
%%%%%%%%%%%%%%%%%%%%%%%%%%%%%%%%%%%%%%%%%%%%%%%%%%%%%%%%%%%%%%%%
\section{Definition of DIM algebra and level $N$ representation}
\label{sec:Def of DIM}

In this section, 
we recall the definition of the DIM algebra and the level $N$ representation. 
For the notations, we follow \cite{FHHSY}. 
The DIM algebra has two parameters $q$ and $t$. 
Let $g(z)$ be the formal series 
\begin{equation}
g(z)\seteq \frac{G^+(z)}{G^-(z)}, \quad 
G^{\pm}(z)\seteq (1-q^{\pm 1}z)(1-t^{\mp 1}z)(1-q^{\mp 1}t^{\pm 1}z). 
\end{equation}
Then this series satisfies $g(z)=g(z^{-1})^{-1}$.

\begin{definition}
Define the algebra $\mathcal{U}$ to be the
unital associative algebra over $\mathbb{Q}(q,t)$ generated by 
 the currents 
$x^\pm(z)=\sum_{n\in \mathbb{Z}}x^\pm_n z^{-n}$,
$\psi^\pm(z)=\sum_{\pm n\in \mathbb{Z}_{\ge0}}\psi^\pm_n z^{-n}$
and the central element $\gamma^{\pm 1/2}$ 
satisfying the defining relations
\begin{align}
&\psi^\pm(z) \psi^\pm(w) = \psi^\pm(w) \psi^\pm(z),
 \quad
 \psi^+(z)\psi^-(w) =
 \dfrac{g(\gamma^{+1} w/z)}{g(\gamma^{-1}w/z)}\psi^-(w)\psi^+(z), \label{eq:def rel of DIM1}
\\
&\psi^+(z)x^\pm(w) = g(\gamma^{\mp 1/2}w/z)^{\mp1} x^\pm(w)\psi^+(z), \\
& \psi^-(z)x^\pm(w) = g(\gamma^{\mp 1/2}z/w)^{\pm1} x^\pm(w)\psi^-(z),
\\
&[x^+(z),x^-(w)]
 =\dfrac{(1-q)(1-1/t)}{1-q/t}
 \big( \delta(\gamma^{-1}z/w) \psi^+(\gamma^{1/2}w)-
 \delta(\gamma z/w) \psi^-(\gamma^{-1/2}w) \big),
\\
&G^{\mp}(z/w)x^\pm(z)x^\pm(w)=G^{\pm}(z/w)x^\pm(w)x^\pm(z) \label{eq:def rel of DIM2}. 
\end{align}
\end{definition}

This algebra $\mathcal{U}$ is an example of the family of
topological Hopf algebras introduced by Ding and Iohara \cite{Ding-Iohara}. 
This family is a sort of generalization 
of the Drinfeld realization of the quantum affine algebras. 
However, 
Miki introduce a deformation of the $W_{1+\infty}$ algebra 
in \cite{Miki:2007}, 
which is the quotient of the algebra $\mathcal{U}$ 
by the Serre-type relation. 
Hence we call the algebra $\mathcal{U}$ 
the Ding-Iohara-Miki algebra (DIM algebra). 
Since the algebra $\mathcal{U}$ has a lot of background, 
there are a lot of other names 
such as quantum toroidal $\mathfrak{gl}_1$ algebra 
\cite{FJMM:2015:Quantum, FJMM:2016:Finite}, 
elliptic Hall algebra \cite{BS:2012:I} and so on.  
This algebra has a Hopf algebra structure. 
The formulas for its coproduct $\Delta$ are 
\begin{align}
&\Delta (\psi^\pm(z))=
 \psi^\pm (\gamma_{(2)}^{\pm 1/2}z)\otimes \psi^\pm (\gamma_{(1)}^{\mp 1/2}z),
\\
&\Delta (x^+(z))=
  x^+(z)\otimes 1+
  \psi^-(\gamma_{(1)}^{1/2}z)\otimes x^+(\gamma_{(1)}z),\\
&\Delta (x^-(z))=
  x^-(\gamma_{(2)}z)\otimes \psi^+(\gamma_{(2)}^{1/2}z)+1 \otimes x^-(z),
\end{align}
and $\Delta(\gamma^{\pm 1/2})=\gamma^{\pm 1/2} \otimes \gamma^{\pm 1/2}$, 
where $\gamma_{(1)}^{\pm 1/2} \seteq \gamma^{\pm 1/2}\otimes 1$
and $\gamma_{(2)}^{\pm 1/2} \seteq 1\otimes \gamma^{\pm 1/2}$.
Since we do not use the antipode and the counit in this paper, 
we omit them.
The DIM algebra $\mathcal{U}$ can be represented by 
the Heisenberg algebra $a_{n}$ ($n\in \mathbb{Z}$) with the relation 
\begin{eqnarray}
[a_n, a_m] = n\frac{1-q^{|n|}}{1-t^{|n|}} \delta_{n+m,0}. 
\end{eqnarray}

\begin{fact}[\cite{FHHSY, FHSSY}]\label{fact:lv. 1 rep of DIM}
Let $u$ be an complex parameter or indeterminate. 
The morphism $\rho_u$ defined as follows is a representation of the 
DIM algebra: 
\begin{align}
&\rho_u(x^+(z))=u\, \eta(z),\quad 
 \rho_u(x^-(z))=u^{-1} \xi(z),\\
&\rho_u(\psi^\pm(z))=\varphi^\pm(z), 
\quad \rho_u(\gamma^{\pm 1/2})=(t/q)^{\pm 1/4}, 
\end{align}
where
\begin{align}
&\eta(z)\seteq
\exp\Big( \sum_{n=1}^{\infty} \dfrac{1-t^{-n}}{n} z^{n} a_{-n} \Big)
\exp\Big(-\sum_{n=1}^{\infty} \dfrac{1-t^{n} }{n} z^{-n} a_n \Big),\\
&\xi(z)\seteq
\exp\Big(-\sum_{n=1}^{\infty} \dfrac{1-t^{-n}}{n}(t/q)^{n/2} z^{n}a_{-n}\Big)
\exp\Big( \sum_{n=1}^{\infty} \dfrac{1-t^{n}}{n} (t/q)^{n/2} z^{-n}a_n \Big),\\
&\varphi_{+}(z)\seteq
\exp\Big(
 -\sum_{n=1}^{\infty} \dfrac{1-t^{n}}{n} (1-t^n q^{-n})(t/q)^{-n/4} z^{-n}a_n
    \Big),
\\
&\varphi_{-}(z)\seteq
\exp\Big(
 \sum_{n=1}^{\infty} \dfrac{1-t^{-n}}{n} (1-t^n q^{-n})(t/q)^{-n/4} z^{n}a_{-n}
    \Big).
\end{align}
\end{fact}

Not that the zero mode $\eta_0$ of $\eta(z)=\sum_n \eta_n z^{-n}$
can be essentially identified with the Macdonald difference operator \cite{Macdonald, SKAO:1995:quantum}. 
By using the coproduct of $\mathcal{U}$, 
we can consider the tensor representation of $\rho_u$. 
For an $N$-tuple of parameters $\vu=(u_1,u_2,\ldots,u_N)$,
define the morphism $\rho_{\vu}^{(N)}$ by
\begin{align}
\rho_{\vu}^{(N)}\seteq
(\rho_{u_1}\otimes\rho_{u_2}\otimes\cdots \otimes\rho_{u_N})
\circ\Delta^{(N)},
\end{align}
where $\Delta^{(N)}$ is inductively defined by 
$\Delta^{(1)}\seteq \mathrm{id} $, $\Delta^{(2)}\seteq \Delta$ and
$\Delta^{(N)}\seteq( \mathrm{id} \otimes \cdots 
\otimes{\rm id}\otimes \Delta)\circ \Delta^{(N-1)}$.
The representation $\rho_{\vu}^{(N)}$ is called 
the level $N$ representation
after the property $\rho_{\vu}^{(N)}(\gamma)=(t/q)^{\frac{N}{2}}$.  
$\rho_{\vu}^{(N)}$ is also called the level $(N,0)$ representation 
or the horizontal representation 
to distinguish another one called 
the level $(0,N)$ representation 
or the vertical representation 
\cite{FFJMM:2011:semiinfinite, FT:2011:Equivariant, awata2012quantum}. 
These representations can be regarded as a sort of duality 
through an automorphism of DIM algebra. 
In the representation $\rho_{\vu}^{(N)}$, 
we write the $i$-th bosons as 
\begin{equation}
a^{(i)}_n \seteq \underbrace{1 \otimes \cdots \otimes 1 \otimes a_n}_i \otimes 1 \otimes \cdots \otimes 1
\end{equation}
for simplicity. 
The generator $X^{(1)}(z)$ in Definition \ref{df:x^i_n} is obtained by 
\begin{equation}
\Xo(z) = \rho^{(N)}_{\vu} (x^{+}(z)). 
\end{equation} 
Note that, in this paper, the parameters $u_i$ are realized by 
the operators $U_i$ and the highest weight vector $\Ket{\vu}$ 
in order to use screening currents. 
%%%%%%%%%%%%%%%%%%%%%%%%%%%%%%%%%%%%%%%%%%%%%%%%%%%%%%%%%%%%%%%%
%%%%%%%%%%%%%%%%%%%%%%%%%%%%%%%%%%%%%%%%%%%%%%%%%%%%%%%%%%%%%%%%
%%%%%%%%%%%%%%%%%%%%%%%%%%%%%%%%%%%%%%%%%%%%%%%%%%%%%%%%%%%%%%%%

%%%%%%%%%%%%%%%%%%%%%%%%%%%%%%%%%%%%%%%%%%%%%%%%%%%%%%%%%%%%%%%%%%
%%%%%%%%%%%%%%%%%%%%%%%%%%%%%%%%%%%%%%%%%%%%%%%%%%%%%%%%%%%%%%%%%%
%%%%%%%%%%%%%%%%%%%%%%%%%%%%%%%%%%%%%%%%%%%%%%%%%%%%%%%%%%%%%%%%%%
%\bibliographystyle{utcaps} 
%\bibliographystyle{myspmpsci} 
%\bibliographystyle{spmpsci}
%\bibliography{myreferences}
%%%%%%%%%%%%%%%%%%%%%%%%%%%%%%%%%%%%%%%%%%%%%%%%%%%%%%%%%%%%%%%%%%
%%%%%%%%%%%%%%%%%%%%%%%%%%%Reference%%%%%%%%%%%%%%%%%%%%%%%%%%%%%%
%%%%%%%%%%%%%%%%%%%%%%%%%%%%%%%%%%%%%%%%%%%%%%%%%%%%%%%%%%%%%%%%%%

\end{document}